\documentclass[aps,prb,reprint,groupedaddress,showpacs,twocolumn]{revtex4}
\usepackage{graphicx}
\usepackage{amsfonts}
\usepackage{amssymb}
\usepackage{amsmath}

\begin{document}

\title{Boundary-field-driven control of discontinuous phase transitions on hyperbolic lattices}

\author{Yoju Lee$^{1}$}
\author{Frank Verstraete$^{1,2}$}
\author{Andrej Gendiar$^{3,}$}\email{andrej.gendiar@savba.sk}

\affiliation{$^{1}$ Faculty of Physics, University of Vienna, Boltzmanngasse 5, A-1090 Vienna, Austria}
\affiliation{$^{2}$ Ghent University, Department of Physics and Astronomy, Krijgslaan 281-S9, B-9000 Gent, Belgium}
\affiliation{$^{3}$ Institute of Physics, Slovak Academy of Sciences, SK-845~11, Bratislava, Slovakia}

\date{\today}

\begin{abstract}
The multistate Potts models on two-dimensional hyperbolic lattices are studied
with respect to various boundary effects. The free energy is numerically calculated
using the Corner Transfer Matrix Renormalization Group method. We analyze phase transitions
of the Potts models in the thermodynamic limit with respect to contracted boundary
layers. A {\it false} phase transition is present even if a couple of the boundary layers
are contracted. Its significance weakens, as the number of the contracted boundary layers
increases, until the correct phase transition (deep inside the bulk) prevails over the
{\it false} one. For this purpose we derive a thermodynamic quantity, the so-called
{\it bulk excess} free energy, which depends on the contracted boundary layers and
memorizes additional boundary effects. In particular, the magnetic field is imposed on
the outermost boundary layer. While the boundary magnetic field does not affect the
second-order phase transition in the bulk if suppressing all the boundary effects on
the hyperbolic lattices, the first-order (discontinuous) phase transition is significantly
sensitive to the boundary magnetic field. Contrary to the phase transition on the Euclidean
lattices, the discontinuous phase transition on the hyperbolic lattices can be continuously
controlled (within a certain temperature coexistence region) by varying the boundary
magnetic field.
\end{abstract}

\maketitle

\section{Introduction}

Negatively curved surfaces have been studied experimentally including lattice dislocations
of solid-state crystals with non-Euclidean properties, e.g., in various magnetic
nanostructures~\cite{exp1,exp2,exp3}, or in materials showing a conical geometry~\cite{conic}.
On the other hand, theoretical interest has been focused on the geometry of anti de Sitter
(AdS) space~\cite{QG1,QG2} and the complex (e.g. neural) networks~\cite{CN1,CN2}.
The nontrivial boundary structure of finite hyperbolic spaces plays an essential role
in the analysis of AdS space by means of entanglement entropy~\cite{tHooft,Susskind,
Takayanagi}. The mutual relations among condensed-matter physics, the general theory of
relativity, and the conformal field theory (CFT) enrich the interdisciplinary research,
such as AdS-CFT correspondence~\cite{AdS-CFT}.

In 2007, we proposed a way to generalize the Corner Transfer Matrix Renormalization
Group (CTMRG) algorithm to a simple hyperbolic surface~\cite{ctmrg_54,ctmrg-tn}.
The current study extends our recent generalization to arbitrary regular hyperbolic
geometry~\cite{ctmrg_pq}, and we focus on the complexity of the boundary effects of
hyperbolic lattice geometry. Because no satisfactory analytic studies are available yet,
we analyze multistate spin models on hyperbolic lattices numerically. In particular,
we focus on phase transitions, which require special treatment because of the
non-negligible boundary effect, as they have been suppressed thus far. The CTMRG algorithm
enables us to calculate the free energy accurately, including all thermodynamic functions
in thermal equilibrium. The free energy approach of the current work has been chosen
because it incorporates all the non-negligible boundary effects of the models on hyperbolic
lattices. The current study is intended to provide complementary and missing information
with respect to the boundary effects performed by Monte Carlo (MC) simulations~\cite{MC1,
MC2,MC3,MC4}.

Hyperbolic lattice geometry is known for exhibiting strong boundary effects, which
prevent the MC simulations from performing sufficiently precise calculations to classify
the phase transitions by means of the free energy. The reason lies in the large fluctuations
coming from the rich boundary structure: the boundary size grows exponentially, as the
diameter (the shortest distance from the lattice center to the boundary) increases
linearly. Hasegawa {\it et al.} studied the Ising model via MC simulations, and they
determined the phase transition by gradually contracting a couple of the boundary
layers~\cite{boundary}. The aim of this paper is to perform a thorough numerical
analysis of the multistate spin models by CTMRG with respect to the boundary layers,
which are gradually contracted from the entire lattice in the thermodynamic limit.

The paper is organized as follows: In Sec.~II we briefly classify an infinite
set of hyperbolic lattices by a pair of two integers $\{p,q\}$. We consider the
multistate Potts model on the negatively curved lattice surfaces. Setion.~III~A is devoted
to the numerical CTMRG analysis of the first- and second-order phase transitions on
square and Bethe lattices. These two lattices are chosen because they are exactly
solvable and can serve as the benchmarks for the future studies. This Section can be
skipped if the reader is familiar with numerical analysis of phase transitions using
thermodynamic functions. Section~III~B contains the core of this work, where
the dependence of the free energy on the gradually contracted layers is shown. We
generalize our study to classify $Q$-state Potts models on regular hyperbolic
lattice geometries. A concise study of the phase transition deeply inside the infinite
bulk is given when contracting the infinite number of boundary layers. A phase-coexistence
region is specified as a temperature interval, within which the first-order
phase transition can be controlled by varying the magnetic field on the boundary layer.
Section~IV is devoted to discussions and concluding remarks.

\section{Lattice geometry and Spin model}

Consider a regular two-dimensional lattice (a curved surface) formed by the tessellation
of congruent polygons with $p$ sides (referred to as the $p$-gons), and the coordination
number $q$ is fixed for the entire lattice. Such a regular lattice can be described by a pair
of integers, $\{p,q\}$, known as the Schl\"{a}fli symbol~\cite{Mosseri}. Depending
on the choice of the two integers $p\geq3$ and $q\geq3$, the Schl\"{a}fli symbol can
describe either Euclidean (flat) or curved (spherical or hyperbolic) lattice geometries.
If the condition $(p-2)(q-2)=4$ holds, only the following three regular Euclidean
two-dimensional lattices exist: the triangular $\{3,6\}$ lattice, the square $\{4,4\}$
lattice, and the honeycomb $\{6,3\}$ lattice. The inequality $(p-2)(q-2) < 4 $ describes
five spherically curved lattices of finite size.
However, if $(p-2)(q-2) > 4$, an infinite set of hyperbolic lattice geometries can be
constructed by tessellation. Each hyperbolic lattice has a
particular constant and negative Gaussian curvature, provided that the size of any
$p$-gon side is identical~\cite{Mosseri}.

Although an arbitrary hyperbolic lattice $\{p,q\}$ of infinite size forms a negatively
curved surface, none of the lattices can be embedded in the three- or finite-dimensional
space. Hence, the Hausdorff dimension is infinite for an arbitrary hyperbolic lattice
$\{p,q\}$, provided that the thermodynamic limit is taken. To visualize the hyperbolic
lattice $\{p,q\}$, it is useful to project each lattice onto a unitary circle;
this mapping is known as the Poincar\'{e} disk representation~\cite{Poincare}. In such a
representation, the congruent $p$-gons (all of the polygons have an identical size and
shape on the entire lattice) are deformed if displayed graphically on the disk. The size of
the $p$-gons decreases toward the circumference of the disk; the circumference represents
the lattice boundary (in the infinity).

%%%%
\begin{figure}[tb]
\centerline{
\includegraphics[width=0.495\textwidth,clip]{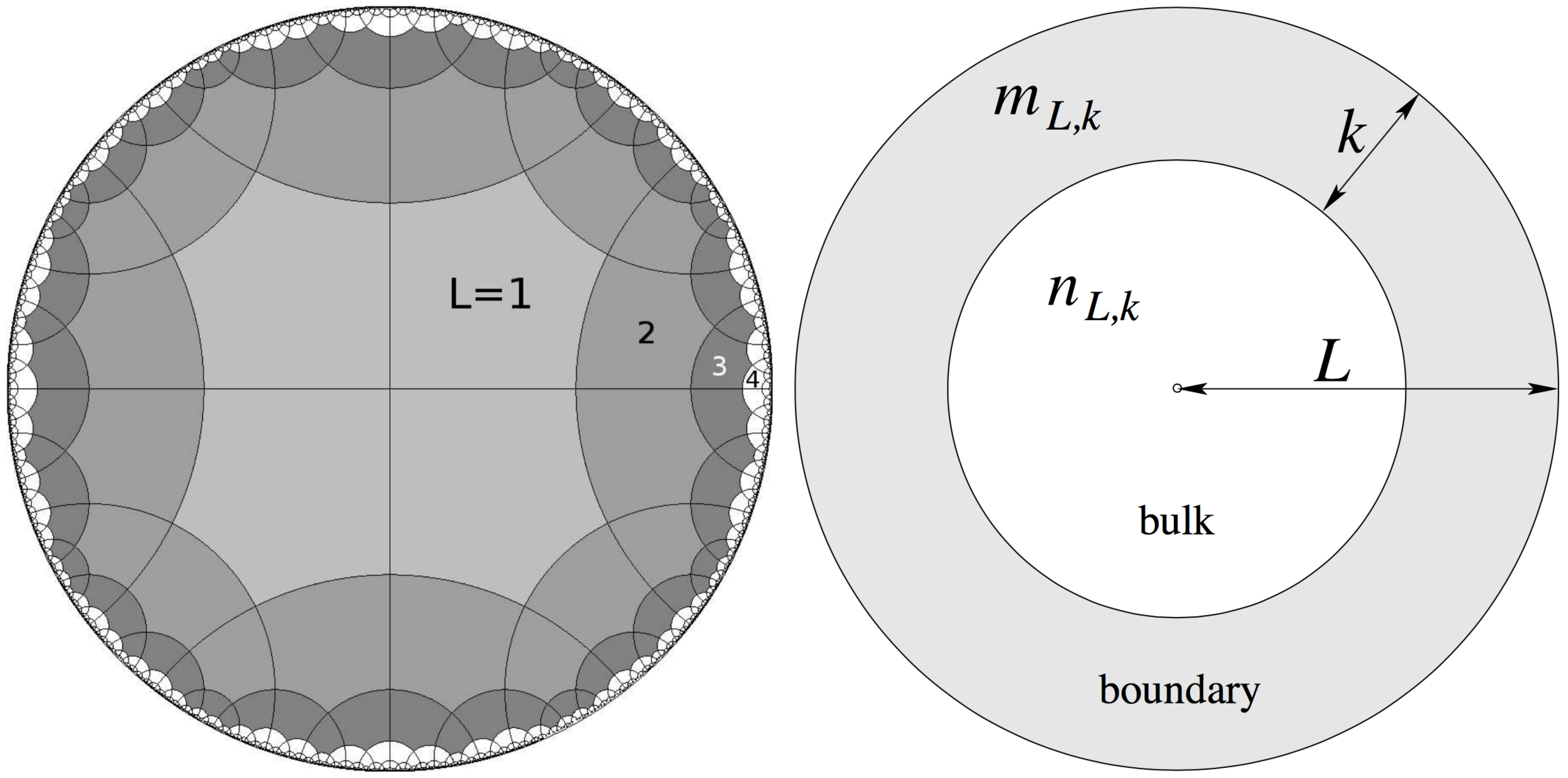}}
\caption{The schematic picture of the hyperbolic lattice $\{5,4\}$ in the Poincar\'{e}
disk representation. The entire hyperbolic lattice is composed of $L$ concentric layers,
where $L$ is referred to as the lattice size (the radius), which enumerates the layers
from the center outward. The inner sublattice, the bulk, has $n_{L,k}$ sites with the
diameter $L-k$ of the layers. The outer layers, the boundary, are enumerated by $k$
and consists of $m_{L,k}$ sites.}
\label{Fig1}
\end{figure}
%%%%

At first we specify the size of the lattice in terms of the radius $L$, which is
well-visible if the hyperbolic lattice is graphically represented in the Poincar\'{e} disk.
Let the integer $L$ denotes the actual radius of the lattice beginning in a central
lattice site. An example of a hyperbolic lattice geometry $\{5,4\}$ is shown in
Fig.~\ref{Fig1} (on the left), where the smallest lattice radius with $L=1$ contains
four regular pentagons. If $L=2$, the total number of the pentagons is $4+20$, etc.
To enhance the $L$-dependence, i.e., the layer structure, the inner layers of the lattice
are depicted by the distinct intensity of the gray color. Now it is obvious how the total
number of the $p$-gons grows exponentially with respect to the linear increase of the
radius $L$ (generalization to any hyperbolic lattice $\{p,q\}$ is straightforward).

\subsection{Boundary layers}

To characterize the boundary layers as a hyperbolic lattice, we ascribe the boundary
layers to an integer variable $k$, which enumerates the number of the outermost layers,
$k=1,2,3,\dots$ (provided that $L>k$). Whenever we refer to the bulk properties of the
(hyperbolic) lattice, one should consider such an innermost sublattice, which has
the radius $L-k$ infinite; or, numerically, $L-k$ is sufficiently large so that all
thermodynamic functions (normalized to the lattice site) has completely converged.
Let $N$ be the total number of the lattice vertices (the sites) for a given diameter $L$.
If $k>0$, the entire lattice can be thought of as a system composed of two parts: the bulk
subsystem containing $L-k$ inner layers and the $k$ outer boundary layers. The total
number of the sites is an integer function of $L$ and is denoted by
${\cal N}^{(L)}_{\{p,q\}}$, which depends on the lattice geometry. The inner $L-k$
layers and the outer $k$ layers, respectively, consist of $n_{L,k}$ and $m_{L,k}$ sites,
i.e., ${\cal N}^{(L)}_{\{p,q\}} = n_{L,k} + m_{L,k}$, as schematically sketched in
Fig.~\ref{Fig1} (on the right).

%%%%
\begin{figure}[tb]
\centerline{
\includegraphics[width=0.495\textwidth,clip]{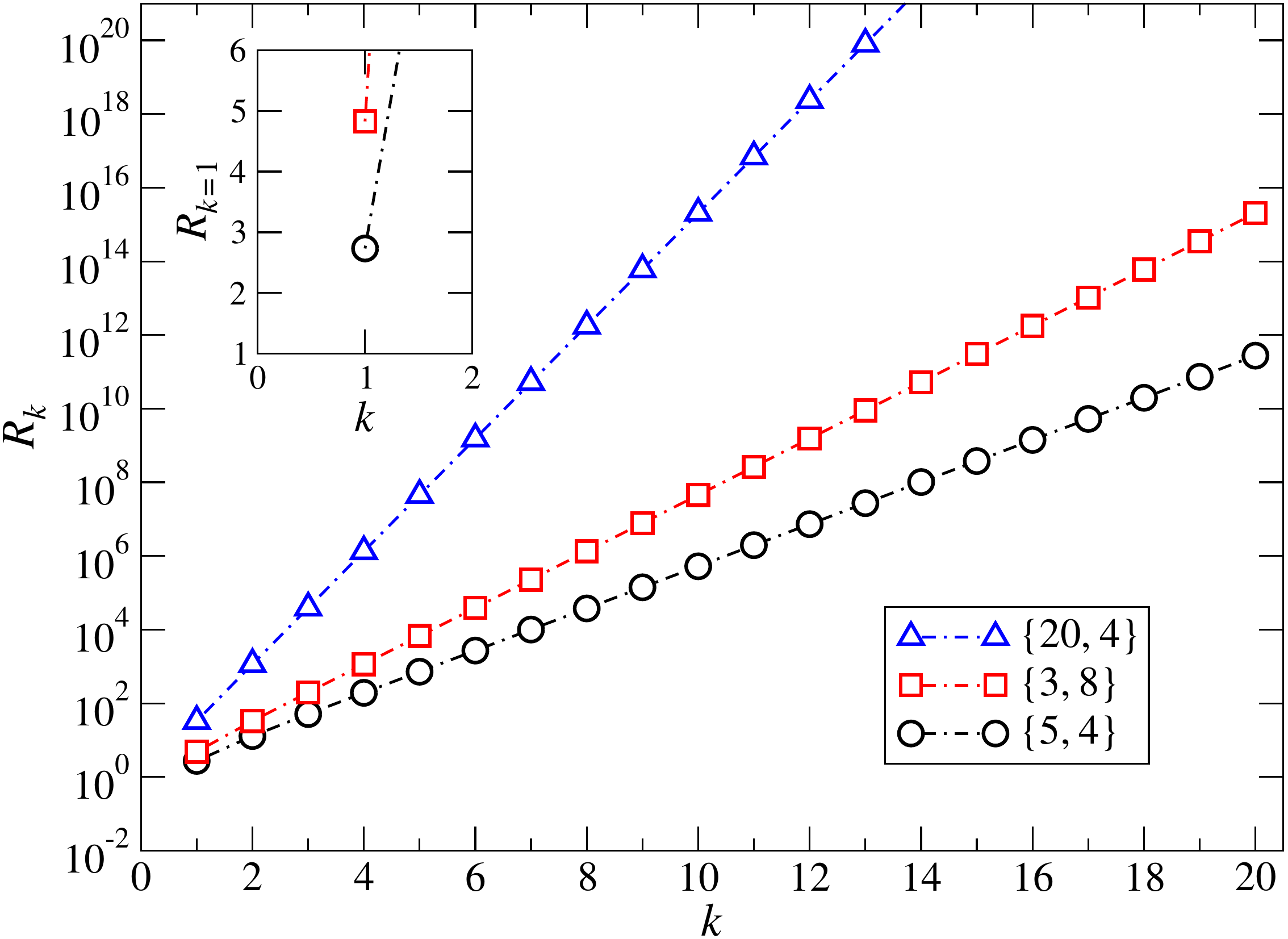}}
 \caption{(Color online) The scaling ratio $R_k$ between the boundary and the bulk
layers is calculated for the three selected hyperbolic lattices $\{20,4\}$, $\{3,8\}$,
and $\{5,4\}$ in the thermodynamic limit ($L\to\infty$). The inset shows the detail
for the outermost layer when $k=1$.}
\label{Fig2}
\end{figure}
%%%%

If we evaluate the ratio between the boundary sites $m_{L,k}$ and the inner bulk sites
$n_{L,k}$, we aim to quantify the significance of the boundary sites, particularly,
\begin{equation}
\nonumber
R_k = \lim\limits_{L\to\infty}\frac{m_{L,k}}{n_{L,k}}\ .
\end{equation}
Recall that $R_{1\leq k < +\infty} = 0$ for the Euclidean lattice geometries. However,
the ratio $R_k$ is always positive for the hyperbolic lattices, as shown in Fig.~\ref{Fig2}.
It is instructive to stress that $R_{k=1}>2$ for the three selected lattices, which means
that the number of the boundary sites is at least twice as large as the inner $L-1$ sites,
even in the thermodynamic limit $L\to\infty$ (the inset). Or, to be more specific,
$R_{1}=2.732$, $4.824$, and $32.971$ on the lattices $\{5,4\}$, $\{3,8\}$, and $\{20,4\}$,
respectively. This clearly implies that an enormous number of sites lies on the
outermost boundary layer, which significantly influence the inner (bulk) properties
that we study in this paper.

\subsection{Potts Model}

Having defined the lattice geometry $\{p,q\}$, now we consider a relevant multistate
spin system on those underlying lattices. Let each lattice vertex (site) be a multistate
spin variable interacting with the $q$ nearest neighboring spins in each $p$-gon. We study
the $Q$-state Potts model~\cite{FYWu} with the nearest-neighbor coupling $J$ (preferably
the ferromagnetic one, $J>0$, to avoid frustration for odd $p$), whose Hamiltonian has the
form
\begin{equation}
 {\cal H} = - J   \sum_{\langle\langle i,j \rangle\rangle} \delta^{~}_{s_i, s_j}
	    - h   \sum_{\langle\langle i   \rangle\rangle} \delta^{~}_{s_i, 0  }
            - h_b \sum_{\langle        i   \rangle_{\rm b}} \delta^{~}_{s_i, 0} \, .
\label{Ham_Potts}
\end{equation}
The Kronecker delta $\delta^{~}_{s_i, s_j}$ acts on the two adjacent (nearest-neighbor)
$Q$-state spin variables $s_i$ and $s_j$ positioned on the vertices $i$ and $j$. A
uniform constant magnetic field $h$ acting on every spin variable $s_i$ is imposed,
whereas an independent magnetic field $h_b$ is applied on the boundary spins only.
Each $Q$-state spin variable $s_i$ possesses $Q$ degrees of freedom, $s_i = 0,1,\dots
,Q-1$, where $Q\geq2$. The summation $\langle\langle i,j \rangle\rangle$ runs over all
the pairs of the nearest-neighboring spins on the entire lattice in the first term.
The summation in the second term, $\langle\langle i \rangle\rangle$, specifies the
interaction of the magnetic field $h$ with the spin site on the lattice in a preferred
spin level $\ell$ (we selected the zeroth spin level, i.e., $\delta^{~}_{s_i,\ell=0}$).
The third term $\langle i \rangle_{\rm b}$ denotes the interaction of the magnetic field $h_b$
with the $m_{L,k=1}$ spin sites located on the outermost boundary layer. When the number
of the spin degrees of freedom is set to be as small as $Q=2$, the model is equivalent
to the Ising model.

For the calculation of the thermodynamic functions, we apply the generalized CTMRG
algorithm~\cite{ctmrg_pq}, which has been proven to be a powerful numerical method
for classical multistate spin lattice models~\cite{ctmrg-tn}. The CTMRG is a classical
counterpart of the Density Matrix Renormalization Group~\cite{White,Uli}. It has
been successfully applied to various spin models on Euclidean and the hyperbolic
lattices~\cite{ctmrg_54,ctmrg_pq,ctmrg_p4,ctmrg_3q,ctmrg_3qn,TPVF54,TPVFp4}. The
algorithm is an efficient and accurate tool to calculate the partition function
${\cal Z}={\rm Tr}\, \exp{(-{\cal H}/k_{\rm B}T)}$, or eventually, the free
energy ${\cal F}=-k_{\rm B}T\ln{\cal Z}$, and the related thermodynamic functions.
Here, $k_{\rm B}$ and $T$, respectively, are the Boltzmann constant and temperature.
In the following we use the dimensionless units, and we consider $k_{\rm B}=1$.
More specific details of the CTMRG algorithm modified for the hyperbolic lattices
are summarized in Refs.~\onlinecite{ctmrg_54,ctmrg_pq}.

To summarize the CTMRG idea in brief, the partition function is considered in terms
of the tensor product of local Boltzmann weights ${\cal W}=\exp{(-{\cal H}_{\{p,q\}}
/k_{\rm B}T)}$ of the $p$-gonal shape, where the Hamiltonian ${\cal H}_{\{p,q\}}$ is
a cyclic spin chain of interacting spins within the $p$-gon for a given $\{p,q\}$
lattice geometry. The algorithm is designed to cover the entire lattice with the
identical Boltzmann weights so as to satisfy the constant coordination number $q$.
The lattice expands iteratively in the algorithm, starting from the $q$ Boltzmann
weight sharing the central site. The lattice grows linearly with the diameter $L$
within the renormalization group procedure until the renormalized corner transfer
tensors converge. The full convergence is associated with the thermodynamic limit,
which is a necessary condition to analyze the phase transitions.

\begin{figure}[tb]
\centerline{
\includegraphics[width=.495\textwidth]{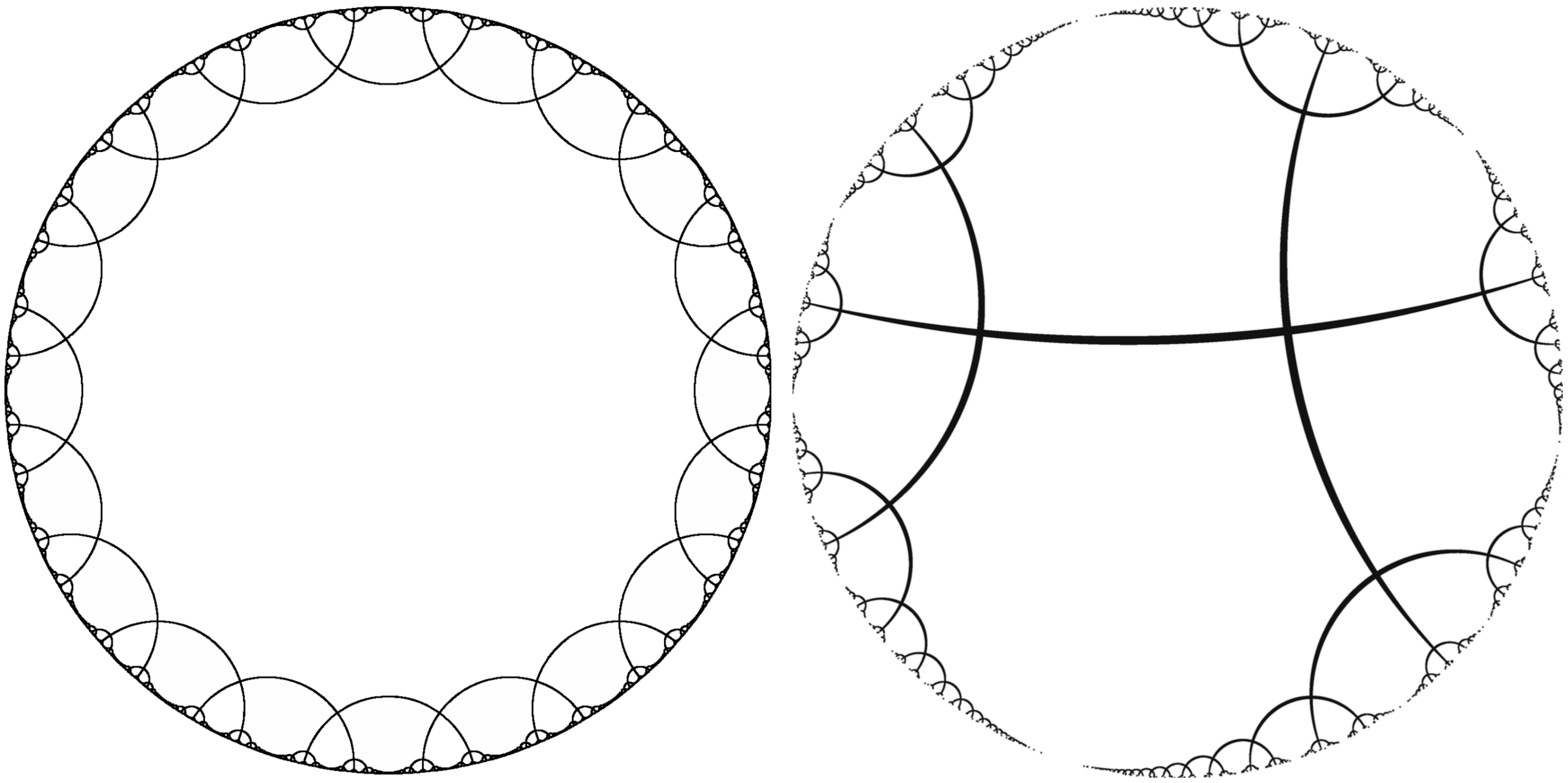}}
  \caption{The hyperbolic lattice geometries $\{20,4\}$ (left) and $\{\infty,4\}$ (right)
are numerically indistinguishable if the bulk properties of the spin models are considered.}
  \label{Fig3}
\end{figure}

The $Q$-state Potts model is exactly solvable~\cite{FYWu} on the square lattice $\{4,4\}$,
resulting in the phase transition temperature $T_{\rm pt}=1/\ln(1+\sqrt{Q})$ for any
$Q\geq2$. In a special case, the two-state Potts model is equivalent to the Ising model,
which has an exact solution on the Bethe lattices~\cite{Baxter} for arbitrary coordination
number $q\geq3$, and the phase-transition temperature is $T_{\rm pt}=1/\ln[q/(q-2)]$.
In the following, we refer to these exact phase transition temperatures, which we use
for comparison with the numerical results. As we have shown earlier~\cite{ctmrg_p4,
ctmrg_pq}, the thermodynamic properties of the spin models in the bulk on the hyperbolic
lattices $\{p\gtrsim15,q\}$ are, in fact, numerically indistinguishable from those on
the Bethe $\{\infty,q\}$ lattices~\cite{Mosseri}, regardless of the coordination number
$q$; cf. Ref.~\onlinecite{ctmrg_p4}. Therefore, we use the hyperbolic lattice $\{20,q\}$
as the Bethe lattice $\{\infty,q\}$. Figure~\ref{Fig3} shows the lattice structure in
the Poincar\'{e} representation of the hyperbolic lattice $\{20,4\}$ (left) being
numerically equivalent to the Bethe lattice $\{\infty,4\}$ (right). The CTMRG
algorithm applied to an arbitrary lattice $\{p,q\}$ is designed to join $q$ corner
transfer tensors around a central site $\sigma_c$; for better visibility,
Fig.~\ref{Fig3} does not display the spin $\sigma_c$ at the center of the unitary
Poincar\'{e} circle for the two lattices~\cite{ctmrg_pq}. Having compared the relative
error $\varepsilon$ at the phase-transition temperature by the CTMRG algorithm with
respect to the analytic result, we obtained high accuracy, $\varepsilon\lesssim
10^{-5}$ for $Q=2$ and $\{20,q\}$.

%%%%
%%%%

\section{Results}

There are two independent calculations in CTMRG to specify the phase transitions.

(i) The simplest one is to calculate the expectation value for the spin (the spontaneous
magnetization)
\begin{equation}
M_{\{p,q\}} = \frac{Q\left[{\rm Tr}\,\left( \delta_{\sigma_{c},0} \rho \right)\right]-1}
                          {Q-1}\, ,
\label{Mg}
\end{equation}
where $\sigma_c$ is the $Q$-state spin variable located in the central site of the
lattice and $\rho$ is a reduced density matrix that is the partial configuration sum of
the product of $q$ corner transfer tensors~\cite{ctmrg_pq}. This approach leads to an
accurate determination of the phase transition temperature in the bulk~\cite{ctmrg_54,
ctmrg_p4}. The effects of the lattice boundaries are strongly suppressed in this type
of calculation. Therefore, the spontaneous magnetization in Eq.~\eqref{Mg} has
been used as a reliable criterion to determine the phase transition in the bulk.
Analogously, the nearest-neighbor (NN) correlation energy,
\begin{equation}
E_{\{p,q\}} = -{\rm Tr}\,\left(J\delta_{\sigma_{c},\sigma_{c+1}} \rho \right)\, ,
\label{En}
\end{equation}
can also be evaluated between two neighboring spins in the center of the lattice. Here,
the phase transition can be extracted from the singularity (a diverging peak) of the
specific heat, which is proportional to taking the numerical derivative of $E_{\{p,q\}}$
with respect to temperature.

(ii) The second, and independent way to find the phase transition is to evaluate the
free energy of the entire system, and the boundary effects contribute significantly to
the free energy. We have recently derived recursive expressions for the free energy
per spin site, ${\cal F}_{\{p,q\}}$ for any multistate spin system including the boundary
magnetic field $h_b$. A detailed analysis of the free energy per site is given in
Ref.~\onlinecite{ctmrg_pq}. Thus, the free energy per spin site (expressed in
the thermodynamic limit) has the form
\begin{equation}
{\cal F}_{\{p,q\}} = - k_{\rm B}T \lim\limits_{L\to\infty}
\frac{\ln{\cal Z}_{\{p,q\}}^{(L)}}{{\cal N}_{\{p,q\}}^{(L)}}\, .
\label{FE}
\end{equation}
We can directly use Eq.~\eqref{FE} since the partition function ${\cal Z}$ can be
calculated to a high numerical accuracy by the CTMRG algorithm, and the total number
of the spin sites ${\cal N}_{\{p,q\}}^{(L)}=n_{L,0}+m_{L,0}$ has an analytic expression
for any $\{p,q\}$ geometry~\cite{ctmrg_pq}. Recall, that the numerical derivatives of
the free energy with respect to temperature $T$ or the magnetic field $h$, yield the
correct thermodynamic functions with the singular behavior at the phase transition.
Below, we refer to the following thermodynamic functions: the normalized entropy per spin,
\begin{equation}
S_{\{p,q\}} = - \frac{\partial {\cal F}_{\{p,q\}}}{\ln(Q)\partial T} \, ,
\label{Se}
\end{equation}
the internal energy,
\begin{equation}
U_{\{p,q\}} = - T^2 \frac{\partial \left( {\cal F}_{\{p,q\}}/T \right)}{\partial T} \, ,
\label{Ui}
\end{equation}
and the specific heat,
\begin{equation}
C_{\{p,q\}} = \frac{\partial_T U_{\{p,q\}}}{\partial T} \, .
\label{Cv}
\end{equation}
The normalization factor $\ln(Q)$ in Eq.~\eqref{Se} is chosen to restrict the entropy
to the interval $0 \leq S_{\{p,q\}} \leq 1$ (note that $S_{\{p,q\}}=0$ at $T=0$
and $S_{\{p,q\}}=1$ at $T\to\infty$ in this work).

\subsection{Continuous versus discontinuous phase transitions}

For tutorial purposes, we select the $2$-state Potts model to represent a second-order
(continuous) phase transition, and the $5$-state Potts model, which is known for exhibiting
a first-order (discontinuous) phase transition on the square lattice $\{4,4\}$. The $Q$-state
Potts models results in the second-order phase transitions if $2\leq Q \leq 4$, whereas the
first-order phase transitions arise if $Q\geq5$ on the two-dimensional Euclidean
lattices~\cite{FYWu}. We first concentrate on the simplest case: no boundary layers are
contracted ($k=0$) and no magnetic field is imposed on the boundary layer ($h_b=0$),
whereas the thermodynamic limit is taken ($L\to\infty$).

\subsubsection*{Ising model on the square and Bethe lattices}

\begin{figure}[tb]
\centerline{
\includegraphics[width=0.495\textwidth,clip]{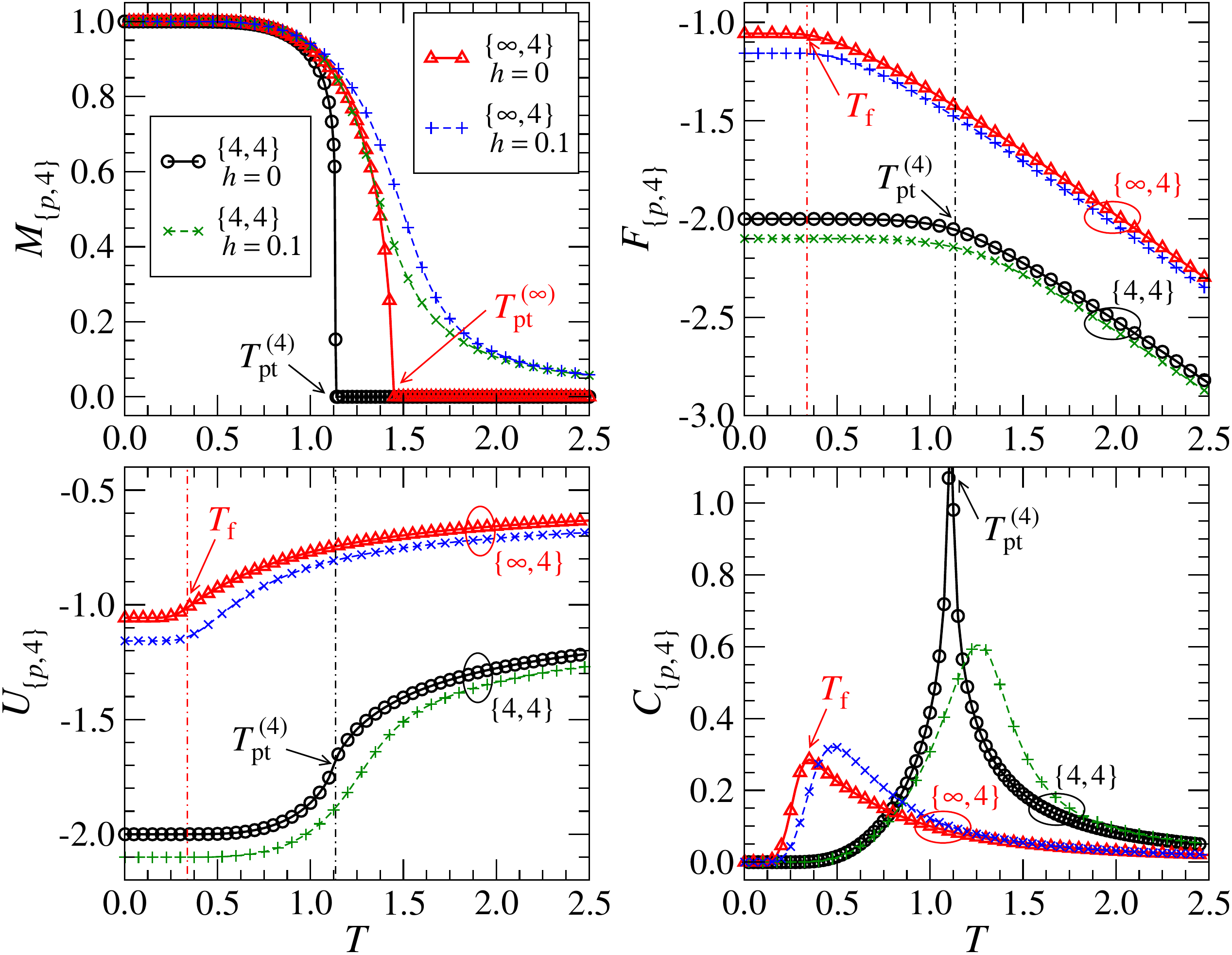}}
 \caption{(Color online) The Ising model: temperature dependence of the spontaneous
magnetization $M_{\{p,4\}}$, the free energy per site ${\cal F}_{\{p,4\}}$, the internal
energy $U_{\{p,4\}}$, and the specific heat $C_{\{p,4\}}$ calculated on the square lattice
$\{4,4\}$ (black circles) and the Bethe lattice $\{\infty,4\}$ (red triangles).
Response of the model to the external magnetic fields $h=0$ (green symbol `$\times$')
and $h=0.1$ (blue symbol `$+$') is shown. The vertical dot-dashed lines help in
identifying the correct phase transition temperature $T^{(p)}_{\rm pt}$ including
the {\it false} one located at $T^{~}_{\rm f}\approx0.339$.}
\label{Fig4}
\end{figure}

First, we investigate the exactly solvable Ising model on the square lattice $\{p=4,4\}$
and the Bethe $\{p\to\infty,4\}$ lattice, and we compare the correctness of the
phase-transition temperature with the exact one. The Ising model exhibits an order-disorder
phase transition at critical temperature $T^{(p=4)}_{\rm pt}=1/\ln(1+\sqrt{2})
\approx1.13459...$ on the square lattice and $T^{(p\to\infty)}_{\rm pt}=1/\ln 2\approx
1.44269...$ on the Bethe lattice. The four typical thermodynamic functions with respect
to temperature for the Ising model are shown in Fig.~\ref{Fig4}. The phase transition
temperature $T^{(p)}_{\rm pt}$ is evident from the typical behavior at zero magnetic
field $h$. The dependence of the four functions at the nonzero field, $h=0.1$, is also
included. The second-order phase transition temperature is well-visible as the
diverging peak in the specific heat [cf. Eq.~\eqref{Cv}], and the free energy does
not exhibit any non-analytic behavior.

%%%%
%%%%

Since the spontaneous magnetization of the Ising model is calculated as the expectation
value of the spin in the center of the lattice (where all the boundary effects are
suppressed), the phase-transition temperature agrees with the analytic result for both
the square and the Bethe lattices. It is located at the temperature $T^{(p)}_{\rm pt}$,
which separates the ordered phase with $M_{\{p,4\}}>0$ from the disordered one, where
$M_{\{p,4\}}=0$ at $h=0$. The calculation of the remaining three thermodynamic functions
${\cal F}_{\{p,4\}}$, $U_{\{p,4\}}$, and $C_{\{p,4\}}$ includes the boundaries. Therefore,
the Ising model on the Bethe lattice shows no evident singularity at $T^{(\infty)}_{\rm pt}$
if the specific heat $C_{\{\infty,4\}}$ for $h=0$ is calculated (cf. $M_{\{\infty,4\}}$).
There is, however, a remarkable (but non-diverging) maximum in $C_{\{\infty,4\}}$ at
significantly lower temperature, which we do not associate with a phase transition.
In the following, we refer to that non-diverging maximum in the specific heat as the
{\it false} phase transition, which is located at temperature $T^{~}_{\rm f}\approx0.339$.
The non-existence of the phase transition on the entire Bethe lattice is not surprising,
as was pointed out in Ref.~\onlinecite{Baxter}. Our numerical analysis using the
free energy, thus, reflects such a feature on the hyperbolic lattice, and it
will be resolved below.

\subsubsection*{5-state Potts model on the square and Bethe lattices}

%%%%
\begin{figure}[tb]
\centerline{
\includegraphics[width=0.495\textwidth,clip]{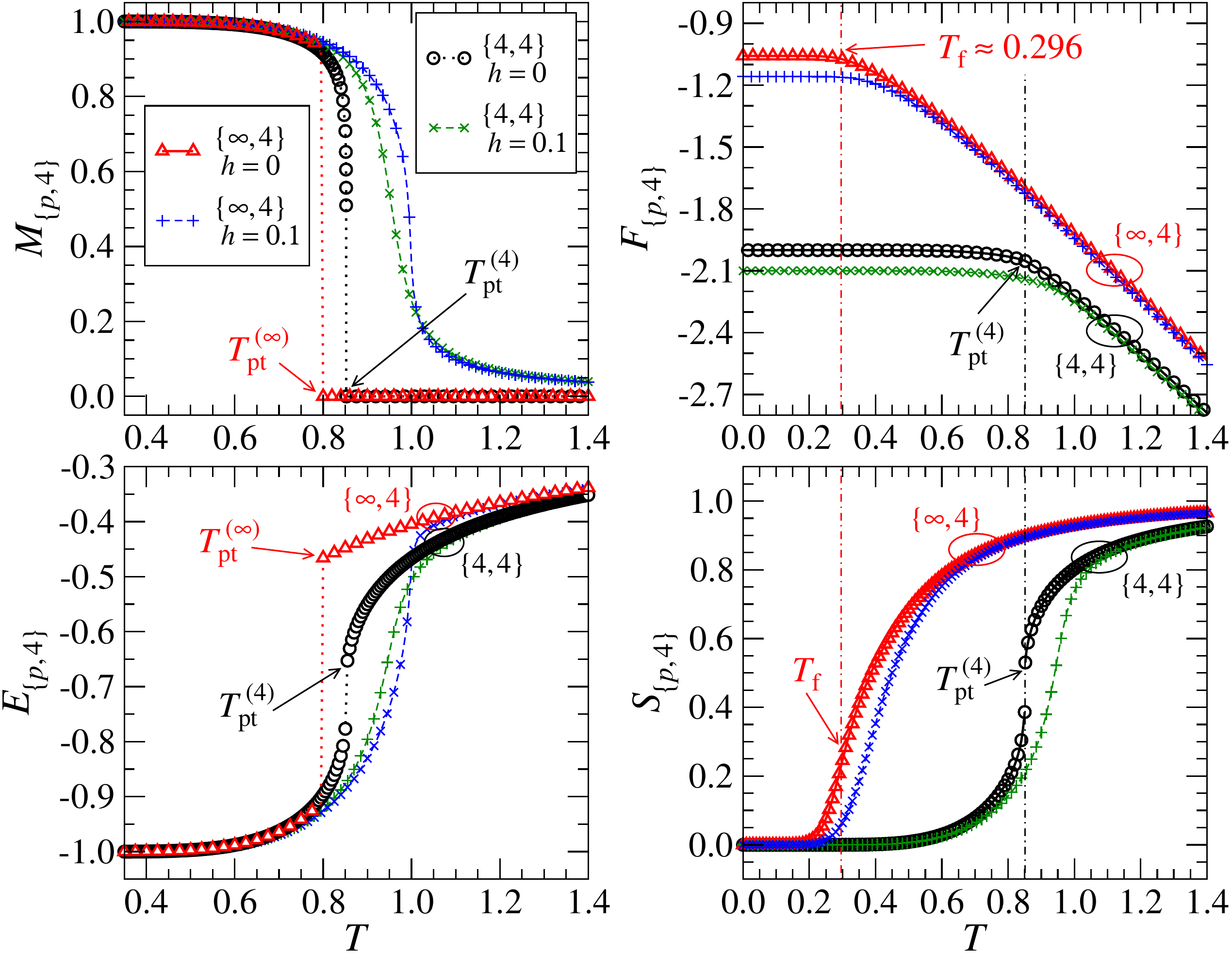}}
 \caption{(Color online) The $5$-state Potts model: the spontaneous magnetization 
$M_{\{p,4\}}$, the free energy per site ${\cal F}_{\{p,4\}}$, the nearest-neighbor
correlation energy $E_{\{p,4\}}$, and the normalized entropy per site $S_{\{p,4\}}$
are plotted with respect to temperature on the square and the Bethe lattices for
$h=0$ and $h=0.1$. The {\it false} phase transition temperature is located at
$T^{~}_{\rm f}\approx0.296$ on the Bethe lattice at $h=0$.}
\label{Fig5}
\end{figure}
%%%%

The first-order phase transition is associated with a sudden (discontinuous) change of
some thermodynamic functions at the phase transition. Specifically, a nonzero latent
heat $\Delta E(T_{\rm pt})>0$ is associated with the discontinuous phase transition at
$T_{\rm pt}$, in particular, $\Delta E \equiv U(T_{\rm pt}^+) - U(T_{\rm pt}^-) \propto
E(T_{\rm pt}^+) - E(T_{\rm pt}^-)$. The nonzero latent heat is the typical signature
of the first-order phase transition, which originates in the nonanalytic behavior of
the free energy at the phase transition. There is a typical kink (crossover) in the
free energy per site, which can be precisely determined by CTMRG after introducing
free and fixed boundary conditions on the Euclidean lattices~\cite{Axelrod}.
To visualize this effect, we plotted a couple of the thermodynamic functions for
the 5-state Potts model in Fig.~\ref{Fig5}. To avoid showing a qualitative similarity
of the internal energy $U_{\{p,4\}}$ and the specific heat $C_{\{p,4\}}$ (as shown in
Fig.~\ref{Fig4}), we replace them by the NN correlation energy $E_{\{p,4\}}$ and the
normalized entropy $S_{\{p,4\}}$, respectively.

The evident discontinuity at the phase transition occurs in $M_{\{4,4\}}$, $E_{\{4,4\}}$,
and $S_{\{4,4\}}$ at $h=0$ [including the less visible kink at $T^{(4)}_{\rm pt}$
in ${\cal F}_{\{4,4\}}$, for details, cf.~Ref.~\onlinecite{Axelrod}].
Obviously, the nonzero latent heat is present, and the method correctly captures
the first-order phase transition temperature, which has an analytic solution~\cite{FYWu}
$T_{\rm pt}^{(4)}=1/\ln(1+\sqrt{5})=0.8515...$ on $\{4,4\}$.

On the other hand, the 5-state Potts model on the Bethe lattice shows different
features. If the boundary effects are neglected, a large discontinuous jump emerges
in the magnetization $M_{\{\infty,4\}}$. There is a jump in the NN correlation energy
$E_{\{\infty,4\}}$, which is proportional to a nonzero latent heat, that is the sign
of the first-order phase transition in the bulk). However, when analyzing the free
energy ${\cal F}_{\{\infty,4\}}$ and the normalized entropy $S_{\{\infty,4\}}$ on
the entire lattice system with the boundary effects, both the free energy and the
entropy, remain continuous (analytic) in the entire temperature interval. Again,
we have localized the {\it false} phase transition, which is located at $T_f\approx
0.296$. The vertical dot-dashed lines in Fig.~\ref{Fig5} serve as guides for the eye
to denote the correct and the {\it false} phase transitions.

\subsubsection*{Basic specifications of discontinuous phase transitions}

%%%%
\begin{figure}[tb]
\centerline{
\includegraphics[width=0.49\textwidth,clip]{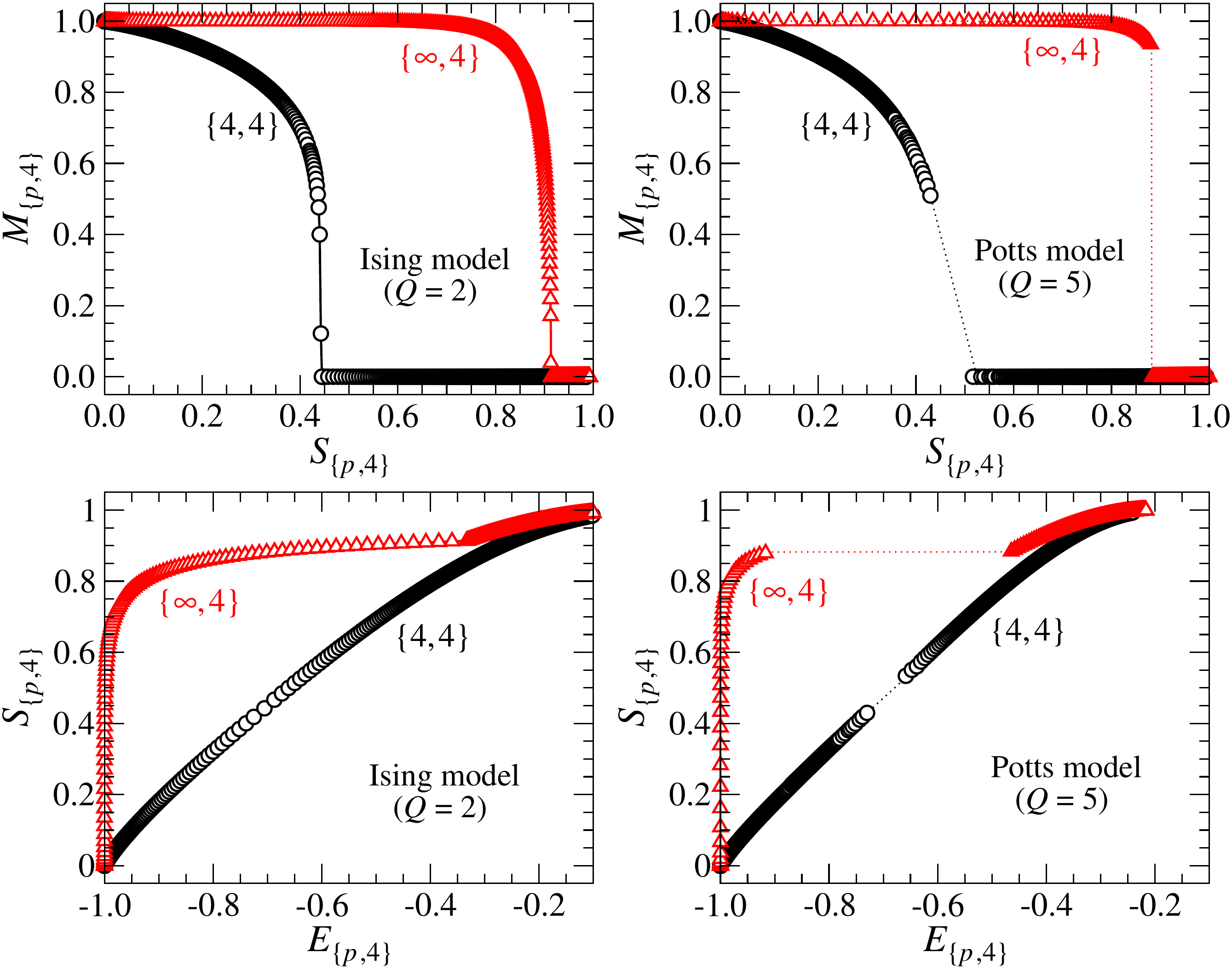}}
 \caption{(Color online) The two top panels show the functional relations between
the magnetization $M$ and the entropy $S$. The two bottom panels show the entropy
$S$ versus the NN correlation energy $E$. The second-order phase transition
is manifested by the continuous behavior of the thermodynamic functions, whereas
the first-order phase transition opens the nonzero gaps in the thermodynamic
functions.}
\label{Fig6}
\end{figure}
%%%%

To emphasize the differences in the thermodynamic properties of the Potts models
on the square and Bethe lattices, we plot mutual functional relations of the
thermodynamic functions in Fig.~\ref{Fig6}. The graphs show the continuous ($Q=2$)
and the discontinuous ($Q=5$) phase transitions from a different viewpoint.
The two top panels depict the functional dependence of the spontaneous magnetization
per site $M_{\{p,4\}}$ on the normalized entropy $S_{\{p,4\}}$ on the square and
the Bethe lattices in the thermodynamic limit. The two bottom graphs show the
normalized entropy per site $S_{\{p,4\}}$ with respect to the NN correlation
function $E_{\{p,4\}}$. All the graphs are evaluated for $J=+1$, $h=0$, and
$h_b=0$ within the temperature interval $0.1\leq T\leq 10$. The Ising model
on the both lattices reproduces the second-order phase transition as the
continuous functions, whereas the first-order phase transition (ascribed to
the case $Q=5$) exhibits the discontinuities in the thermodynamic functions.

%%%%
\begin{figure}[tb]
\centerline{
\includegraphics[width=0.49\textwidth,clip]{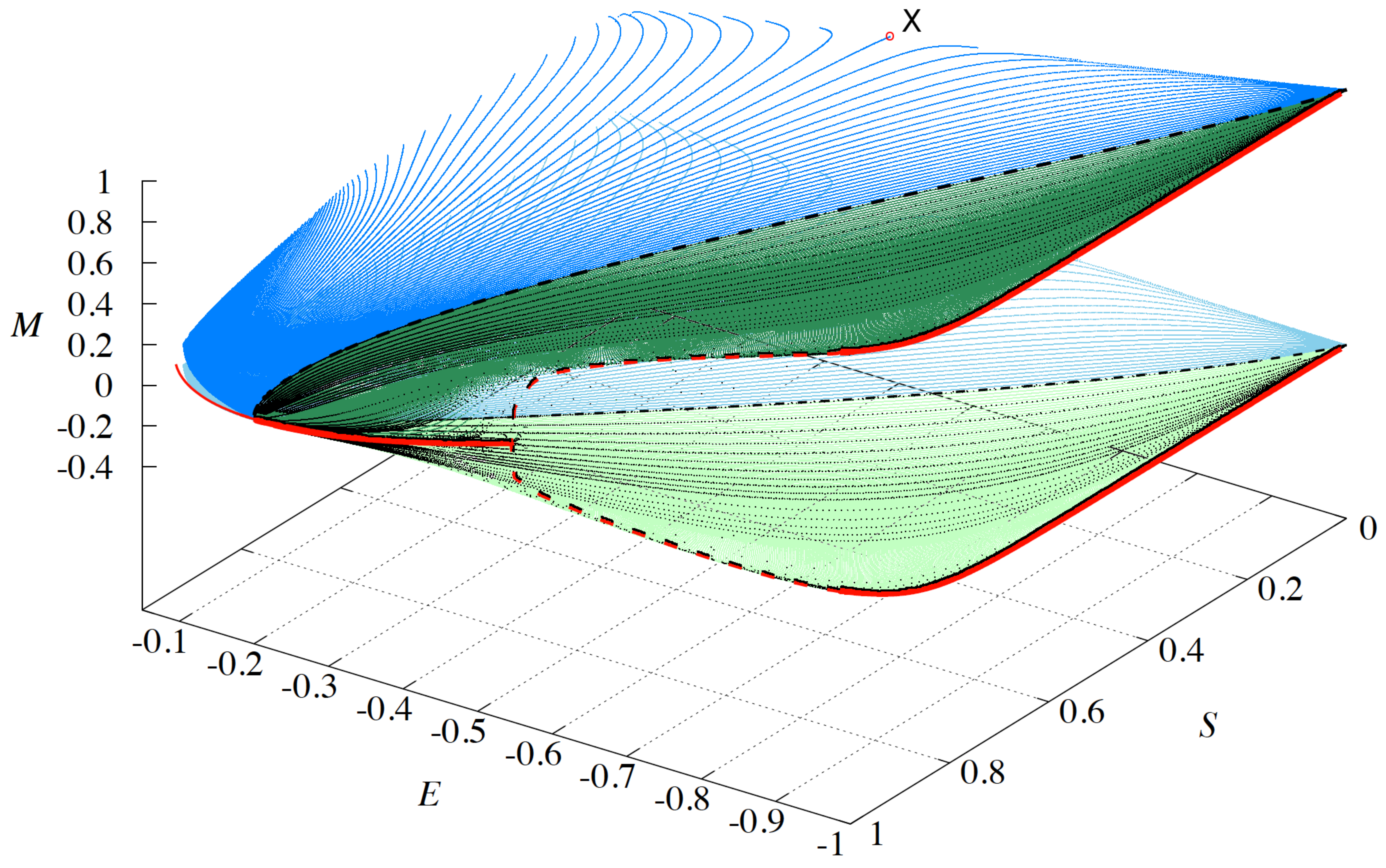}}
 \caption{(Color online) The convex set of $\{E,S,M\}$ for the 5-state Potts model on
the Bethe lattice $\{\infty,4\}$ forms a two-dimensional surface of a shell-like object.
Each thick or thin surface curves correspond to the particular flows of $\{E,S,M\}$
with respect to the varying temperature $T$ for given fixed values $J$ and $h$. The
data shown are calculated for $h_b=k=0$ in the thermodynamic limit ($L\to\infty$).}
\label{Fig7}
\end{figure}
%%%%

If one is interested in understanding the global phase structure of various systems,
a convex set~\cite{Cirac,Mazziotti} is a useful tool for such a demonstration. The
convex set is usually constructed for three typical thermodynamic functions to reveal
the details of various types of the phase transitions if plotting their mutual functional
dependences. Since the convex sets have been constructed neither for the first-order
phase transitions nor for the non-Euclidean lattices, we plot Fig.~\ref{Fig7} to display
the convex set $\{E,S,M\}$ for the 5-state Potts model on the Bethe lattice. We chose
the NN correlation function $E\equiv E_{\{\infty,4\}}$, the normalized entropy $S\equiv
S_{\{\infty,4\}}$, and the spontaneous magnetization $M\equiv M_{\{\infty,4\}}$ to
construct $\{E,S,M\}$ after having scanned sufficiently wide intervals of the
parameters $T$, $J$, and $h$. In particular, we calculated the thermodynamic functions
in the intervals of the temperatures $0.1 \leq T \leq 10$, the coupling constants
$-2 \leq J \leq +2$, and the uniform magnetic field $-10 \leq h \leq +10$. The
constructed convex set has to be understood as the functional set
$\{E,S,M\}\equiv\{E(T,J,h),S(T,J,h),M(T,J,h)\}$.
The convex set $\{E,S,M\}$ of the 5-state Potts model forms a two-dimensional,
]semi-open, and curved surface with a shell-like structure. The surfaces of the upper
and the lower shells, respectively, correspond to $h>0$ and $h<0$. It is worth
mentioning that the present convex set differs remarkably from those obtained for the
models on the Euclidean surfaces. For instance, if the Ising model on the $\{4,4\}$
lattice is considered (not shown), the convex set covers a smaller surface region and
is identical to that evaluated in Ref.~\onlinecite{Zauner}.
The full thick red curves (bordering the front edges) correspond to the case $J=1$ and $h=0$
when the entire temperature interval is scanned. The red dashed curves connecting the
full curves serve only as guides for the eyes. The dashed curves denote a discontinuous
jump (it is used to connect the two full thick red lines), and the discontinuity cannot
by reached by any $T$, $J$, and $h$. The unique jump is, therefore, explicitly shown
in the two panels in Fig.~\ref{Fig6} (on the right). The discontinuity can be
asymptotically approached for small magnetic fields $h$. The point, at which the full
and the two dashed red curves meet is a kind of {\it repeller} point at $\{E\approx0.467,
S\approx0.822,M=0\}$. No discontinuity of that kind has been reported yet.
The thin black dashed curves refer to the zero NN spin coupling ($J=0$); they
divide the top (bottom) shell into two regions (the green one with $J>0$ and the blue
one with $J<0$). The sign of the spontaneous magnetization $M$ coincides with the sign
of the magnetic field $h$.
Each thin curve on the surface of $\{E,S,M\}$ is parameterized by the temperature
$T$ for the fixed values $J$ and $h$. When restricting to the upper shell only, there
are three important {\it attractors} (fixed points) in the graph: ${\cal A}_{\infty}$,
${\cal A}_{1}$, and ${\cal A}_{2}$. The attractor ${\cal A}_{\infty}$ is connected
either with ${\cal A}_{1}$ or ${\cal A}_{2}$ while varying temperature $0\leq T <
+\infty$. The attractor ${\cal A}_{\infty}$ is positioned at $\{E=-0.2,S=1,M=0\}$ at
$T\to\infty$, and it can be reached for arbitrary $J$ and $h$. The two distinct attractors
${\cal A}_1$ and ${\cal A}_2$ can be accessed at $T=0$: the first one at $\{E=-1,S=0,
M=+1\}$ and the second one at $\{E=0,S=0, M=0\}$ (not shown due to strong frustration
effects caused by the competing antiferromagnetic interaction $J<0$ and the uniform
magnetic field, $h\neq0$; the numerical data are not reliable if the frustration
becomes strong, i.e. if $E_{\{p,q\}}>-0.1$). The two attractors ${\cal A}_1$
and ${\cal A}_2$ located at $T=0$, which is equivalent to $S=0$ on the graph, are
mutually inaccessible by varying the temperature $T$. This is due to a bifurcation
point $X$ that separates them (cf. Fig.~\ref{Fig7}). The bifurcation point is
located at $\{E\approx-0.4,S=0,M\approx0.62\}$ on the upper shell.
The top and the bottom shells do not show a mirror symmetry with respect to the plane
at $M=0$. The bottom shell surface (for negative $M$) is less shallow if compared
to the upper one. This is due to the Potts model spin levels, for which the spontaneous
magnetization of the 5-state Potts model exhibits five-fold degeneracy of the
free energy below the phase transition. The free energy minima at $T<T_{\rm pt}$ are
associated with $Q=5$ projections of the spontaneous magnetization in Eq.~\eqref{Mg}
being polarized at $T=0$ either to the spin level with the magnetization $M=+1$ or
to the four identical levels with $M=-\frac{1}{4}$ (cf.~Ref.~\onlinecite{Axelrod}),
provided that the spontaneous symmetry-breaking has occurred in the thermodynamic
limit [because the magnetic field $h$ is in effect up to the spin level $\ell=0$,
the term $h\delta_{\sigma_i,\ell=0}$ is used in Hamiltonian~\eqref{Ham_Potts}].
Apart from the atypical discontinuous region denoted by the thick dashed red curves,
there is another interesting feature: both the top and the bottom shells asymptotically
reach the zero entropy limit $S\to0$ as $T\to0$ for the entire spectrum of $E$, which
is not accessible on the Euclidean lattice geometry~\cite{Zauner}, where the entropy
of the bifurcation point is nonzero (i.e., $S>0.5$ for the Ising model on $\{4,4\}$).

\subsection{Analysis of the boundary effects}

As we have mentioned earlier, the boundary of the hyperbolic lattices affect the
bulk properties, such as the suppression of the phase transition. At the same time,
a {\it false} phase transition appears, which has an obvious origin in the rich
boundary structure. Therefore, we propose a thermodynamic function for any
underlying geometry $\{p,q\}$
\begin{equation}
{\cal B}^{(k)}_{\{p,q\}} = {\cal F}_{\{p,q\}} - {\cal F}^{\ast(k)}_{\{p,q\}}\, ,
\label{bulk}
\end{equation}
where ${\cal F}_{\{p,q\}}$ is the free energy per site [cf. Eq.~\eqref{FE}] of the
entire system $L$ in the thermodynamic limit and ${\cal F}^{\ast(k)}_{\{p,q\}}$ is
the free energy of the $k$ outermost boundary layers. To avoid confusions, we stress
that ${\cal B}^{(k)}_{\{p,q\}}$ is not identical to an isolated free energy of the
$L-k$ layers. Instead, it is meant to describe a thermodynamic function, that
represents {\it bulk excess} free energy, ${\cal B}^{(k)}_{\{p,q\}}$. It contains
all the contributions coming from the $k$ contracted boundary layers (via
normalization factors within the bulk, which depends recursively on each of the $k$
boundary factors~\cite{ctmrg_pq}). Such a specific definition of the {\it bulk excess}
free energy in Eq.~\eqref{bulk} has been chosen as the only numerically feasible
quantity, that can determine the correct phase transition, including the
boundary effects. (For completeness, ${\cal B}^{(k=0)}_{\{p,q\}} \equiv {\cal F}_{\{p,q\}}$
if $k=0$ only.)

\subsubsection*{Ising model on Bethe lattice for $k\to\infty$}

First, the {\it bulk excess} free energy is tested in two asymptotic regimes: $k=0$ and
$k\to\infty$. We focus on the phase transition analysis in the Ising model ($Q=2$) on the
Bethe lattice $\{\infty,4\}$. If $k\to\infty$, we consider the case when both the number
of bulk layers, $L-k$, and the number of boundary layers, $k$, are infinite. To achieve
this, one can, e.g., take $k=\frac{L}{2}$ and perform the thermodynamic limit $L\to\infty$.
Whenever we refer to the two asymptotic cases ($L\to\infty$ and $k\to\infty$), we mean
sufficiently large (finite) values $L$ and $k$, for which any further increase of either of
them has not affected the thermodynamic functions in Eqs.~\eqref{Mg}-\eqref{Cv} at all. Since
these thermodynamic functions are normalized per spin site, they always have to converge
completely. Typically, we set $L = 6000$ for $k = 3000$ (which is the case of $L,k\to\infty$).
We have checked that the numerical results remain identical if $k$ is chosen arbitrarily,
regardless of the order in which the two limits $k$ and $L$ are taken.

%%%%
\begin{figure}[tb]
\centerline{
\includegraphics[width=0.495\textwidth,clip]{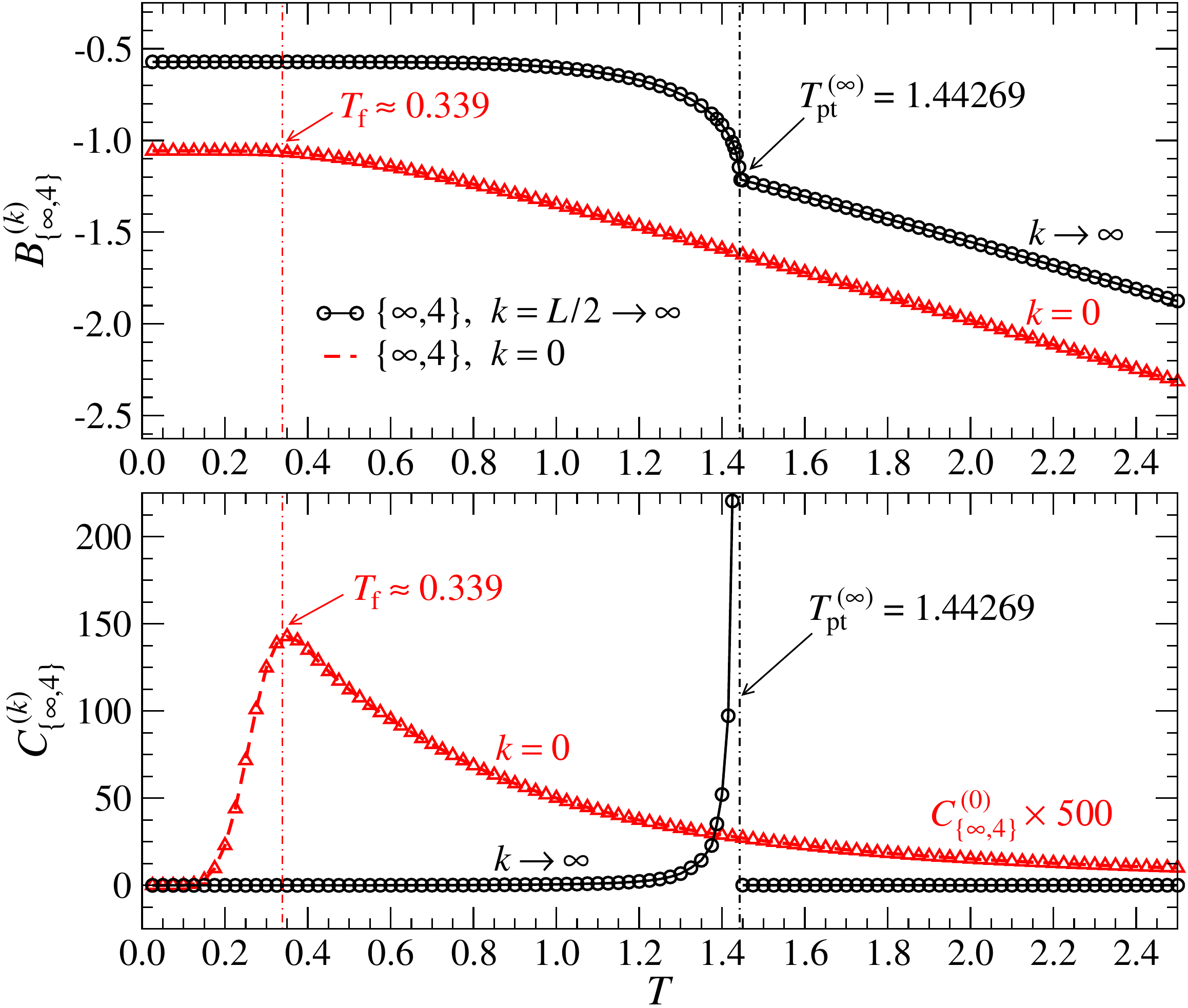}}
 \caption{(Color online) The {\it bulk excess} free energy (the upper graph) and the
`specific heat' (the lower graph) with respect to temperature in the thermodynamic limit
calculated for the Ising model on the Bethe lattice when $h=0$, regardless of $h_b$.
The two cases are shown: (1) $k=0$, $L\to\infty$ (the circles), i.e. no contracted boundary
layers, and (2) $k\to\infty$, $L\to\infty$ (the triangles), i.e. infinitely many boundary
layers are contracted while the inner lattice size remains infinite, too. The two dot-dashed
vertical lines denote the correct and the {\it false} phase transition temperatures.}
\label{Fig8}
\end{figure}
%%%%

The upper graph of Fig.~\ref{Fig8} shows the {\it bulk excess} free energy per site,
${\cal B}^{(k)}_{\{\infty,4\}}$, for the Ising model on the Bethe lattice when $k=0$
and $k\to\infty$. There is an evident singularity in the {\it bulk excess} free energy
in the correct phase transition, $T^{(\infty)}_{\rm pt}=1/\ln 2$. The lower graph shows
the second derivative of the {\it bulk excess} free energy with respect to temperature,
\begin{equation}
{\cal C}^{(k)}_{\{\infty,4\}}=-T\frac{\partial^2{\cal B}^{(k)}_{\{\infty,4\}}}{\partial T^2}
\end{equation}
for both $k=0$ and $k\to\infty$, which we refer to as the `specific heat' in the following
(where the term "specific heat" is true only if $k=0$). The singularity of the `specific heat'
for $k\to\infty$ occurs exactly at the phase-transition temperature $T^{(\infty)}_{\rm pt}
=1/\ln 2$, as is analytically derived in the bulk of the Bethe lattice~\cite{Baxter}.
The singularity is identical to that in the spontaneous magnetization in Fig.~\ref{Fig4}.
If an arbitrary boundary magnetic field $h_b$ is imposed on the boundary layer, both
the phase transition temperature $T^{(\infty)}_{\rm pt}$ and the {\it bulk excess} free
energy ${\cal B}^{(\infty)}_{\{\infty,4\}}$ remain unaffected by setting any $h_b\neq0$
(not shown).
An additional remark is inevitable: the square lattice results in ${\cal B}^{(k)}
_{\{4,4\}} \equiv {\cal F}_{\{4,4\}}$, regardless of whether we took $k=0$ or
$k\to\infty$ in the thermodynamic limit. Moreover, if an arbitrary boundary magnetic
field $h_b$ is imposed, all thermodynamic functions in the bulk are unaffected by it.
This is true for any spin model exhibiting a second-order phase transition on Euclidean
lattices, since the boundaries are negligible [$R_k = 0$ (for any $k\geq0$)].

\subsubsection*{Ising model on Bethe lattice for $0\leq k <\infty$}

%%%%
\begin{figure}[tb]
\centerline{
\includegraphics[width=0.49\textwidth,clip]{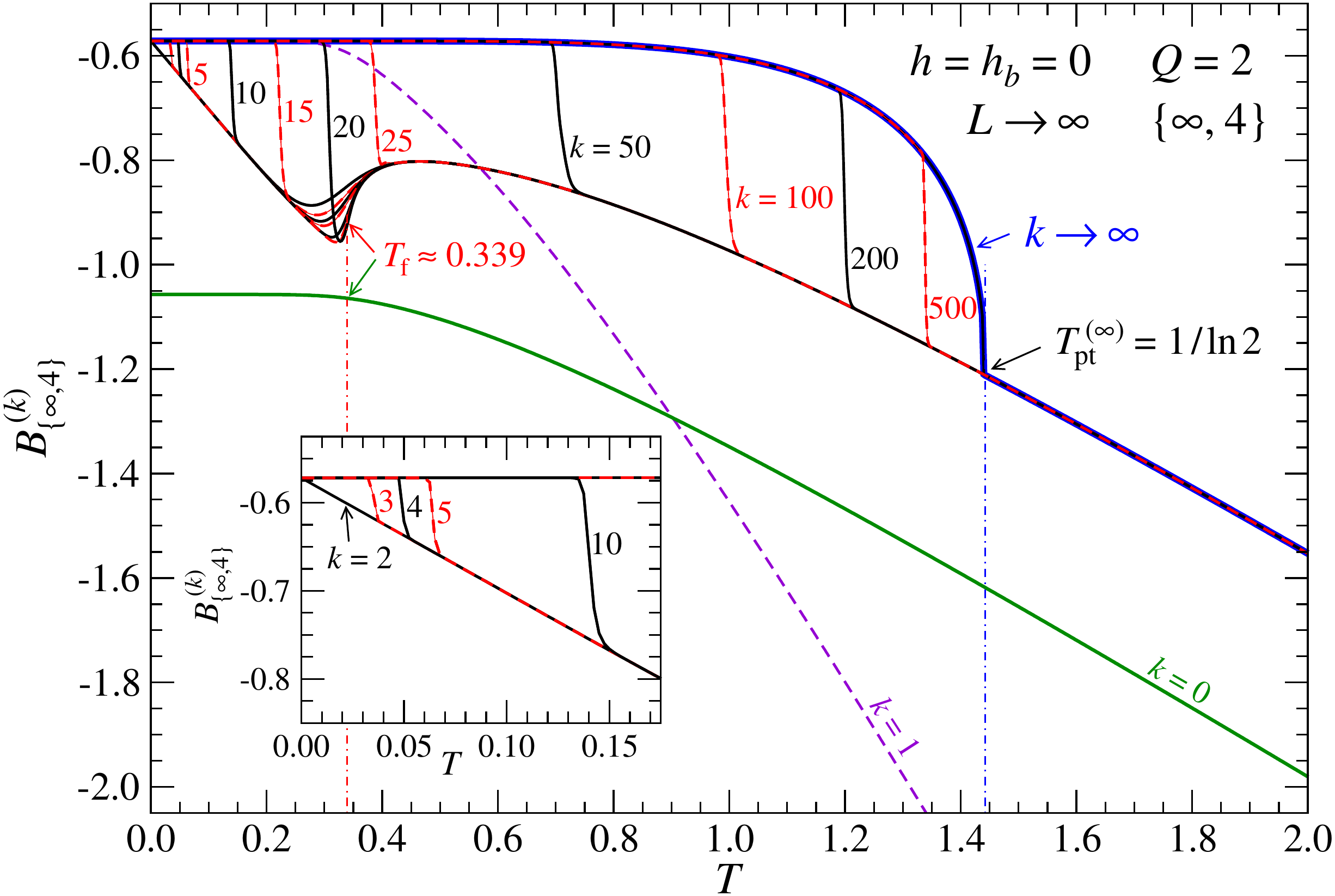}}
 \caption{(Color online) Temperature dependence of the {\it bulk excess} free energy
per site for the Ising model on the Bethe lattice for a wide range of the contracted
boundary layers $k$. Notice the specific region located around the {\it false} phase
transition $T_{\rm f}$, where the {\it bulk excess} free energy grows, is denoted by
the dot-dashed arrows.}
\label{Fig9}
\end{figure}
%%%%

\begin{figure}[tb]
\centerline{
\includegraphics[width=0.49\textwidth,clip]{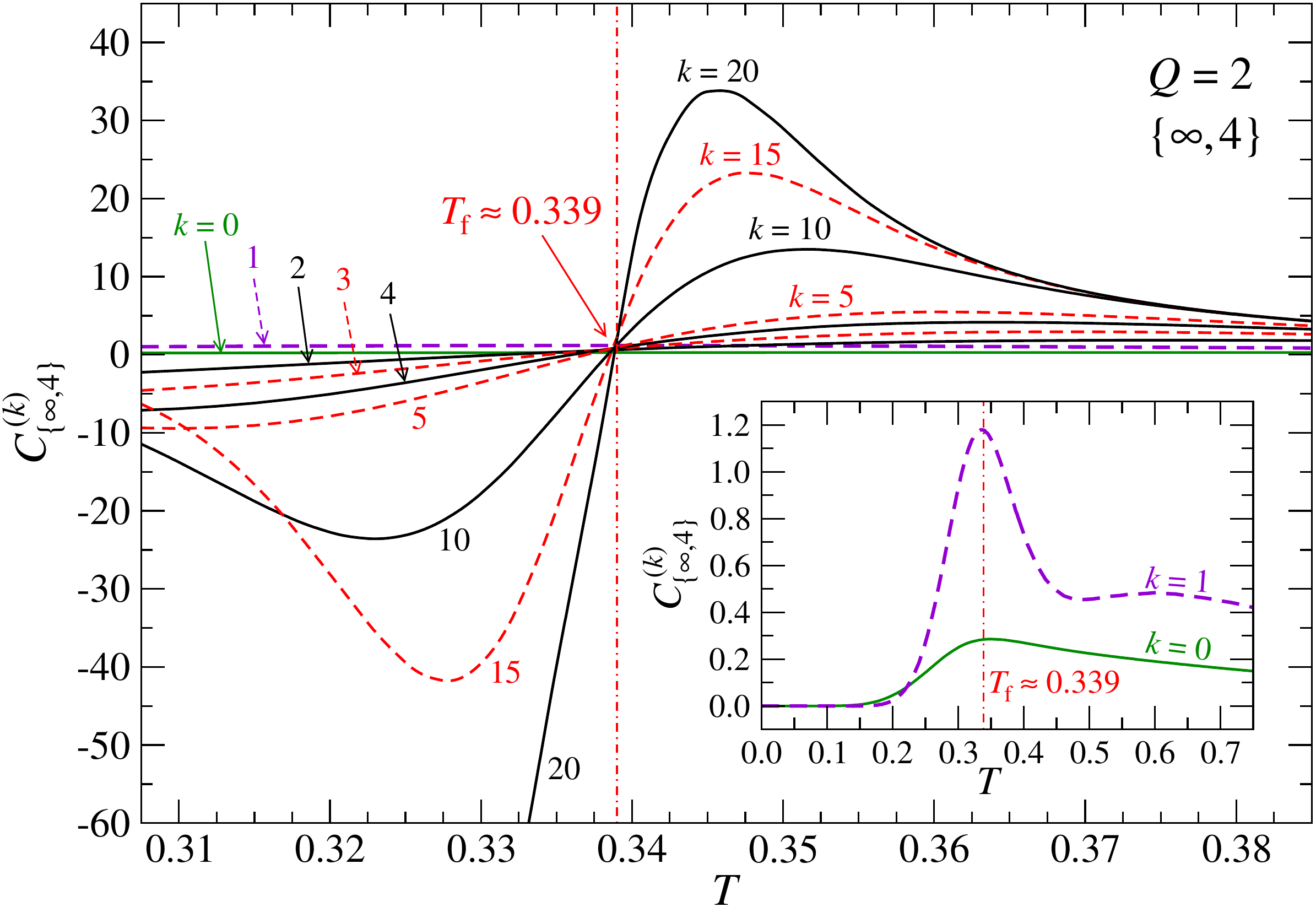}}
 \caption{(Color online) The specific heat vs temperature exhibits the unique behavior
around the {\it false} phase transition temperature $T_{\rm f}\approx 0.339$ after the
boundary layers $k=0,1,2,3,\dots,20$ are gradually contracted. The inset shows the
maximum for $k=0$ and $k=1$.}
\label{Fig10}
\end{figure}

From the previous analysis we came to understand the importance of contracting an infinite
number of the outermost layers in order to detect the correct phase transition in the
bulk by means of ${\cal B}^{(\infty)}_{\{\infty,4\}}$. Now, we analyze {\it bulk excess}
free energy if we gradually contract the individual boundary layers (indexed by $k$).
The aim is to reveal how the {\it false} phase transition at $k=0$ gets suppressed, and
the correct phase transition appears at $k\to\infty$. As an instructive example, the
Ising model on the Bethe lattice is again considered.
Figure~\ref{Fig9} shows the $k$-dependence of the {\it bulk excess} free energy with
respect to temperature for $h=h_b=0$. For comparison, the two limiting cases $k=0$ and
$k\to\infty$ are also plotted. As the number of the contracted boundary layers $k$
increases, the {\it bulk excess} free energy for $k=1,2,3,\dots,\infty$ undergoes a
specific regime. We also observe dramatic changes of ${\cal B}^{(k)}_{\{\infty,4\}}$
toward the asymptotics ${\cal B}^{(k\to\infty)}_{\{\infty,4\}}$ (shown by the thick
blue line in the limit $k\to\infty$ in accord with the upper graph of Fig.~\ref{Fig8}).
We stress an abrupt (continuous) decrease of the {\it bulk excess} free energy for
$k=3,4,...,500$. The higher $k$ is, the closer the abrupt decrease in the asymptotic
correct phase transition occurs.
The {\it false} phase transition at $T_{\rm f}\approx0.339$ (depicted in Figs.~\ref{Fig4}
and \ref{Fig8}) is denoted by a thin vertical dot-dashed line and is relevant for
$0 \leq k \lesssim 20$ in this particular case. It is clear that the {\it bulk excess}
free energy remains significantly influenced by the lattice boundaries, and the {\it false}
phase transition $T_{\rm f}$ prevails over the correct phase transition $T_{\rm pt}$ for
$k\lesssim 20$.
If $k=1$, there is a surprisingly distinct asymptotic behavior of the {\it bulk excess}
free energy ${\cal B}^{(1)}_{\{\infty,4\}}$ at $T>T_{\rm f}$ compared to $k>1$. This is
also manifested in the fact that after contracting the outermost boundary layer, the {\it
bulk excess} free energy changes dramatically, despite the fact that the majority of
lattice sites were disregarded ($R_1>1$). For illustration purposes, the total number of
outermost boundary spins $m_{\infty,k=1}$ is approximately 33 times larger on the lattice
geometry $\{20,4\}$ than the spins inside, cf. Fig.~\ref{Fig2}. (Recall that the regular
Bethe lattice $\{\infty,4\}$ gives $R_{k=1}\to\infty$.) 
Now, we focus on the interval $0 \leq k \lesssim 20$, where $T_{\rm f}$ dominates.
The {\it bulk excess} free energy increases locally around $T_{\rm f}$, which is not
allowed for the free energy in Eq.~\eqref{FE}. There is an inflection point, which
is exactly positioned at $T=T_{\rm t}$. To investigate this feature in detail, we
calculate the `specific heat' for each $k$ ($0 \leq k \lesssim 20$) separately, as
shown in Fig.~\ref{Fig10}.

%%%%
%%%%

Let us point out that the `specific heat' for $k=1$ exhibits a maximum at the identical
$T_{\rm f}\approx0.339$, as plotted in the inset of Fig.~\ref{Fig10}. As $k$ consequently
increases ($k=2,3,4,\dots,\sim20$), there is no maximum anymore. However, the crossover
in ${\cal C}^{(2 \leq k \lesssim 20)}_{\{\infty,4\}}=0$ is a testament to the uniqueness
of $T_{\rm f}\approx0.339$ [notice the inflection point in ${\cal B}^{(2 \leq k
\lesssim 20)}_{\{\infty,4\}}$ which causes the crossover].

For $k \gtrsim 20$, the {\it false} phase transition is completely suppressed, and
the abrupt decrease of ${\cal B}^{(k \gtrsim 20)}_{\{\infty,4\}}$ accompanies the
singularity, which converges to correct phase transition in the bulk, $T_{\rm pt}^{(k
\to\infty)}=1/\ln2$. We reached the complete numerical convergence of the {\it bulk
excess} free energy after contracting $k\geq 2\times10^3$ layers.

\subsubsection*{Ising model on hyperbolic $\{p,q\}$ lattices for $k\to\infty$}

\begin{figure}[tb]
\centerline{
\includegraphics[width=0.49\textwidth,clip]{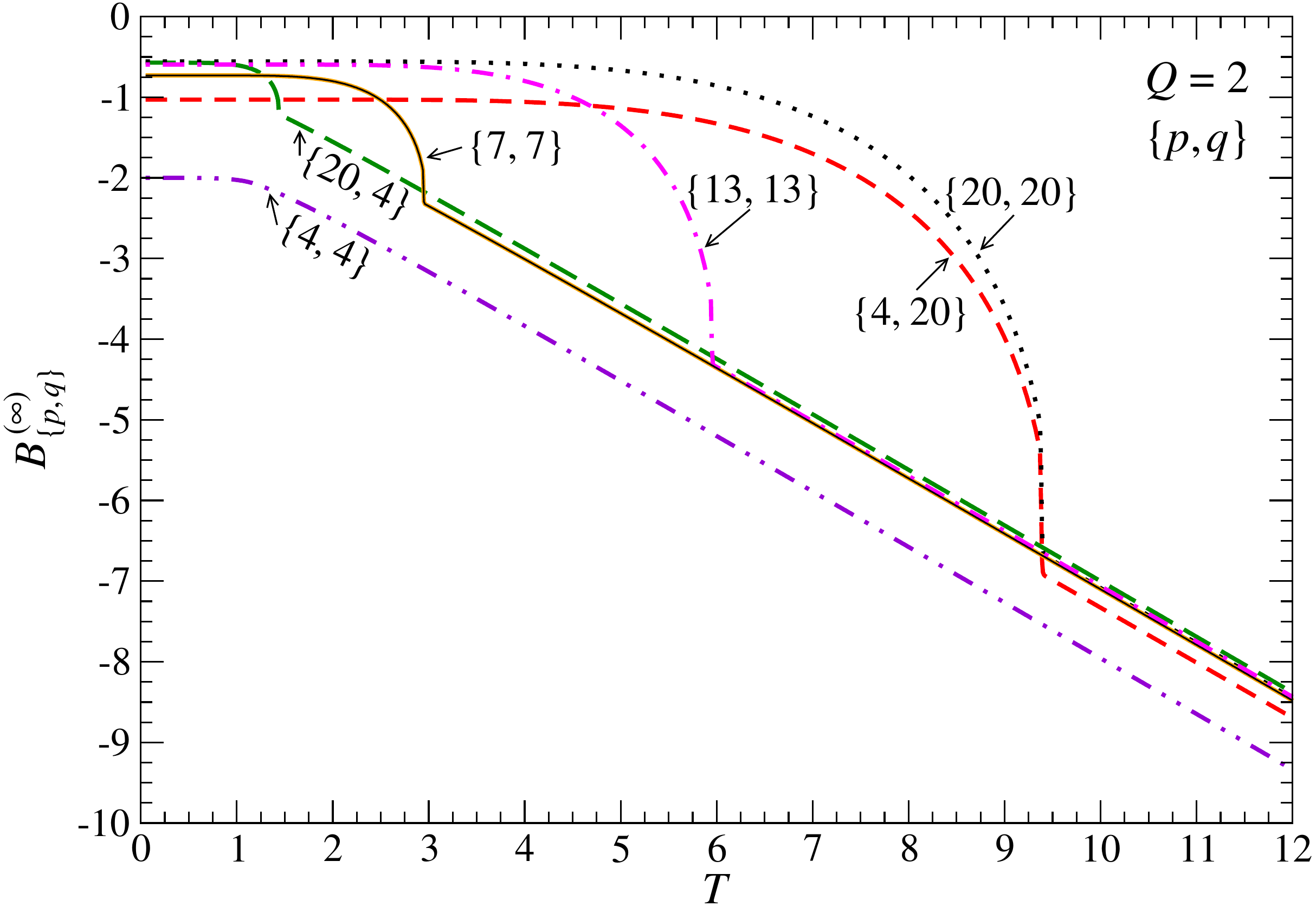}}
 \caption{(Color online) The {\it bulk excess} free energy of the Ising model on the selected
set of the hyperbolic lattices $\{p,q\}$ when taking both the limits $L\to\infty$
and $k\to\infty$ for $h=0$. The {\it bulk excess} free energy remains unchanged for arbitrarily
chosen $h_b \neq 0$.}
\label{Fig11}
\end{figure}

An analogous dependence of the {\it bulk excess} free energy ${\cal B}^{(k)}_{\{p,q\}}$
on $k$ was confirmed numerically on different lattice geometries for the Ising model.
Figure~\ref{Fig11} depicts ${\cal B}^{(k)}_{\{p,q\}}$ for the Ising model on the
several representative lattice geometries in the thermodynamic limit when $k\to\infty$.
Imposing $h\neq0$ on the Ising model on various hyperbolic $\{p,q\}$ shifts the {\it bulk
excess} free energy as in Fig.~\ref{Fig4}. The profile of the {\it bulk excess} free
energy at $k\to\infty$ does not depend on $h_b$ when the Ising model is considered,
provided that $h=0$ and $L\to\infty$.
At sufficiently high temperatures, $T \gg T_{\rm pt}^{(\infty)}$, the slopes of
${\cal B}^{(k)}_{\{p,q\}}$ became identical for any $k\geq0$. In other words, if we
evaluate the normalized entropy per site in Eq.~\eqref{Se}, then $S=1$ for $T\to\infty$.
This is a numerical confirmation that the entropy of the $Q$-state spin models has to
saturate at $\ln Q$ at high temperatures regardless of $k$ and the lattice
geometry~\cite{ctmrg_pq}.

%%%%

\subsubsection*{5-state Potts model on Bethe lattice for $k\to\infty$}

The $Q$-state Potts model with $Q>2$ exhibits a discontinuous (first-order) phase
transition on hyperbolic lattices $\{p,q\}$. We have chosen the $5$-state Potts model
on the Bethe lattice as a typical example. Discontinuous phase transitions
characteristically exhibit a small temperature region, where two (or more) phases
can coexist. For instance, a simple order-disorder phase transition occurs for the
$Q$-state Potts models, where the discontinuous phase transition $T_{\rm pt}$ can
be located. In such a case, the free energy exhibits a kink at $T_{\rm pt}$ on the
Euclidean lattices~\cite{Axelrod}. The phase transition is not critical since the
correlation length does not diverge. The kink in the free energy means that taking
its first derivative with respect to temperature results in a discontinuous function,
a nonzero gap in the internal energy, at the phase transition. The gap $\Delta E
(T_{\rm pt})>0$ is proportional to the latent heat, which is a clear sign of the
first-order phase transition.
We do not study the $k$ dependence of the {\it bulk excess} free energy for the 5-state
Potts model, as it is analogous to the 2-state Potts model. Instead, we focus on the
special features of the first-order phase transition on the hyperbolic geometries.

\begin{figure}[tb]
\centerline{
\includegraphics[width=0.49\textwidth,clip]{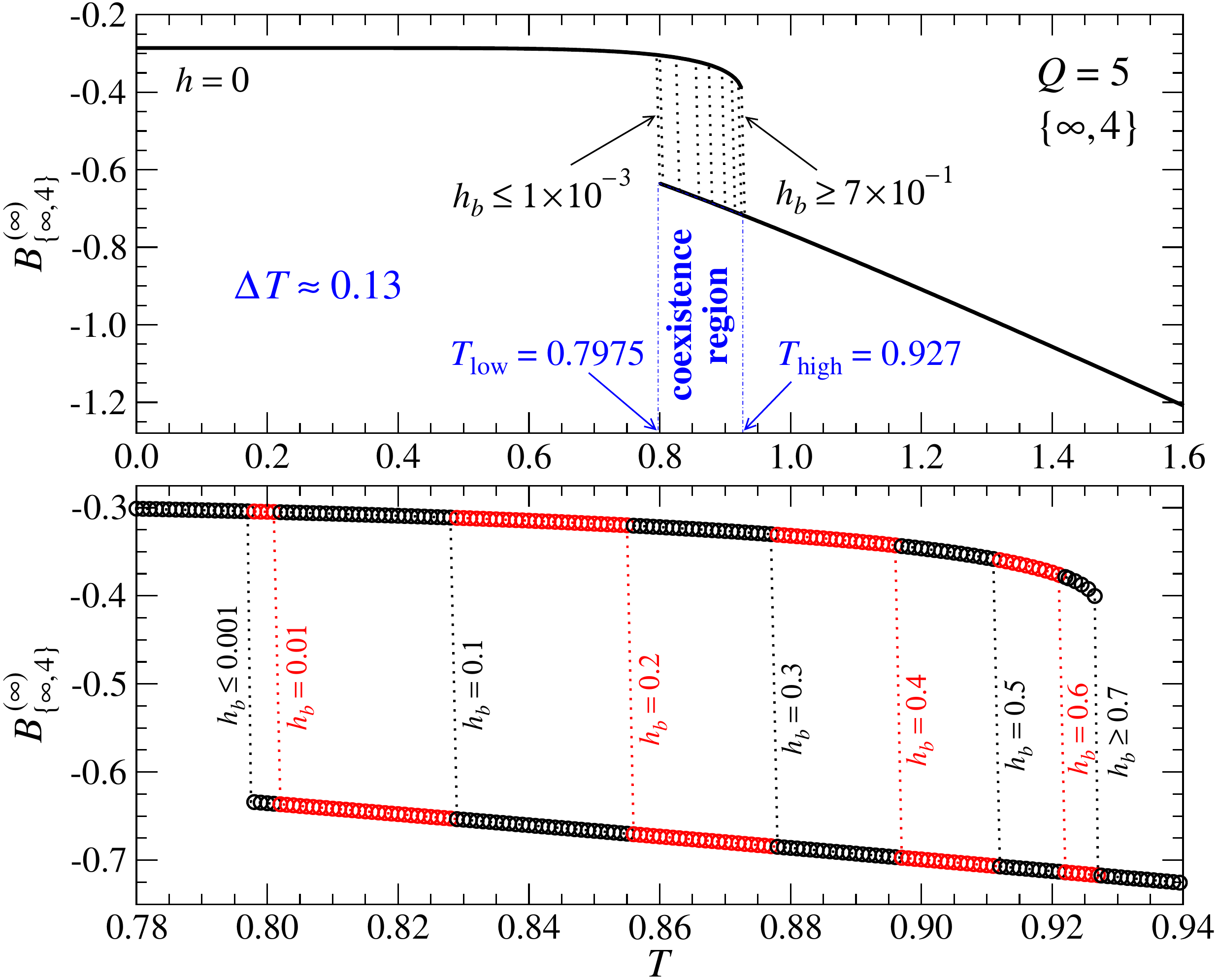}}
 \caption{(Color online) The temperature dependence of the {\it bulk excess} free energy
for the $5$-state Potts model on the Bethe lattice $\{\infty,4\}$ after contracting the
infinite number of the boundary layers ($k\to\infty$). The phase transitions are shown as
the discontinuities in ${\cal B}^{(\infty)}_{\{\infty,4\}}$ for the particular boundary
fields $h_b$. The discontinuities are marked by the dotted vertical lines. The graphs
were constructed for $L\to\infty$, $k\to\infty$, and $h=0$.}
\label{Fig12}
\end{figure}

If a model shows a discontinuous phase transition on the hyperbolic lattices, no 
kink in the free energy or the {\it bulk excess} free energy is found. The top graph in
Fig.~\ref{Fig12} shows the temperature dependence of the {\it bulk excess} free energy
for the $5$-state Potts model on the Bethe lattice when $L\to\infty$ and $k\to\infty$
at zero magnetic field $h$. We have considered a large variety of boundary conditions,
which are determined by the boundary magnetic field $h_b$. We found out that the singular
behavior of the {\it bulk excess} free energy for $k\to\infty$, yields many phase transitions
$T_{\rm pt}(h_b)$, which are sensitive to the imposed magnetic field $h_b$. Within the
interval $-\infty < h_b < +\infty$, we locate a phase-coexistence region on the
finite-temperature interval $\Delta T$. The singular behavior in
${\cal B}^{(\infty)}_{\{\infty,4\}}$ appears at $T_{\rm pt}$ and is accompanied by
a discontinuous jump (denoted by the dotted line). The bottom graph zooms-in on the
relevant part of the top graph, namely, the phase-coexistence region, and it shows,
for which $h_b$ the phase transition $T_{\rm pt}(h_b)$ occurs. (We remark that there
were no such discontinuities of the {\it bulk excess} free energy for
the Ising model: the abrupt decrease with respect to $k$ was always continuous.)

%%%%
%%%%

\begin{figure}[tb]
\centerline{
\includegraphics[width=0.49\textwidth,clip]{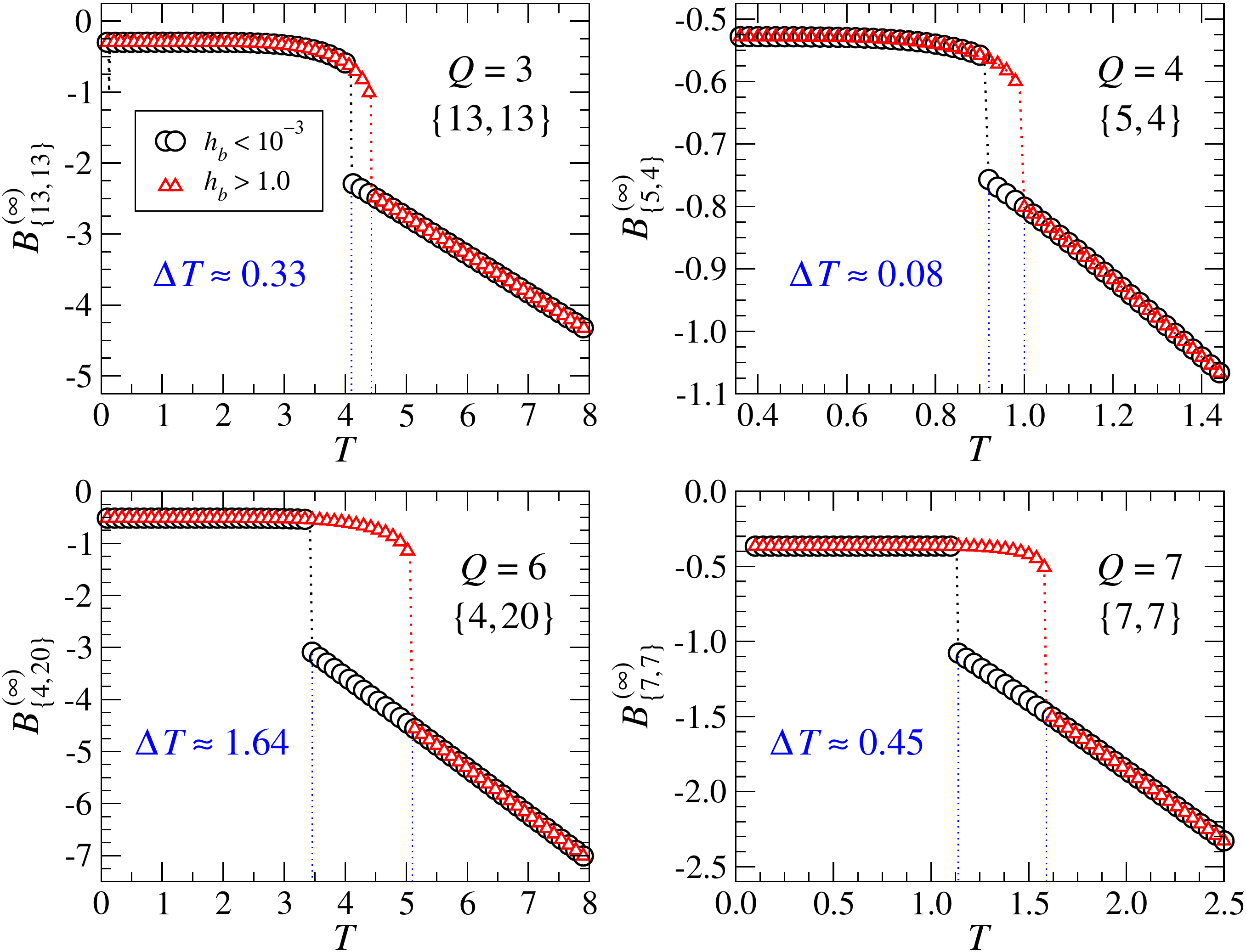}}
 \caption{(Color online) The temperature dependence of the {\it bulk excess} free energy
for the $Q$-state Potts model for the selected Potts states $Q\geq3$ and lattices $\{p,q\}$.
The vertical dotted lines border the discontinuous phase coexistence regions $\Delta T$ for
$L\to\infty$, $k\to\infty$, and $h=0$.}
\label{Fig13}
\end{figure}

Thus, the most emerging feature is the discontinuity of the {\it bulk excess} free
energy while tuning the magnetic fields in the interval $10^{-3}\leq h_b\leq 0.7$.
The finite-temperature interval determining the phase-coexistence region $\Delta T =
T_{\rm high} - T_{\rm low}$ has a lower bound at $T_{\rm low}\approx 0.7975$, which
is stable for any boundary magnetic fields $h_b \lesssim 10^{-3}$ (including the
negative fields), and an upper bound $T_{\rm high}\approx0.927$, which is saturated
at $h_b \gtrsim 0.7$. 
Figure~\ref{Fig13} extends our findings to more examples of the $Q$-state Potts
models on various hyperbolic lattices. Since the Hausdorff dimension of the hyperbolic
geometries $\{p,q\}$ is infinite for arbitrary $(p-2)(q-2)>4$, the first-order phase
transition has to be realized for $Q\geq3$. This is related to the fact that the $Q$-state
Potts model on the three-dimensional cubic lattice~\cite{FYWu,3DPotts} exhibits
first-order phase transitions if $Q\geq3$, whereas the Potts model on the two-dimensional
Euclidean lattices shows this feature if $Q\geq5$.

%%%%
%%%%

In general, the phase-coexistence region $\Delta T$ has a lower bound at $T_{\rm low}
(h_b \ll 0.1)$ and a finite upper bound at $T_{\rm high}(h_b \gg 0.1)$. The width of
the phase coexistence region $\Delta T>0$ is primarily proportional to the lattice
coordination number $q$ and the number of spin states $Q$, whereas there is a weak
dependence on the $p$ (the larger the $q$ and/or $Q$, the wider $\Delta T$). We,
therefore, conjecture that the first-order phase transition temperature $T_{\rm pt}$
is continuously adjustable within the interval $\Delta T$ by varying the boundary
magnetic field $h_b$, which is achieved for the Potts models with $Q\geq3$ and for
an arbitrary hyperbolic geometry $\{p,q\}$.

\section{Discussion and Remarks}
	
The reason for studying the complex behavior of the multistate spin models on the
hyperbolic geometries lies in the potential to mimic the thermodynamic properties of
various neural structures, social networks, anti de Sitter spaces, etc. We have
introduced a way of specifying the novel function in the thermodynamic limit, the
so-called {\it bulk excess} free energy per spin site, which is correlated with the
number of contracted boundary spin layers of the entire lattice. At the same time,
the {\it bulk excess} free energy remains sensitive to the magnetic field imposed
on the outermost boundary layer. Recall that the phase transitions on the Euclidean
lattices are not influenced by choosing any type of the boundary conditions in the
thermodynamic limit.
We have studied the boundary response to the bulk properties and the phase transitions
in the classical multistate spin models on various hyperbolic lattices in the
thermodynamic limit. We applied CTMRG to the $Q$-state Potts models and calculated the
free energy per site with high accuracy. The Potts model has been selected for
its availability to describe both second- and first-order phase transitions.

With respect to our results, there is no phase transition present on any hyperbolic
lattice if the entire lattice in the thermodynamic limit is considered, provided that
no boundary layers were contracted. To be more specific, there is a {\it false} phase
transition induced by the huge boundary structure, which is strong enough to prevail
over the correct phase transition observed deep inside the system if suppressing the
boundaries. However, if a sufficient number of boundary layers are contracted (while
keeping the infinite number of inner layers), the correct phase transition can
appear in the bulk (known from the analytic solutions). Our results agrees with the
only available analytic solution of spin models on Bethe lattices.

We have conjectured that the multistate spin systems on hyperbolic lattices, which
exhibit a continuous (second-order) phase transition, exhibit a firm phase-transition
point irrespective of the imposed boundary magnetic field $h_b$. Note that such
spin models on hyperbolic lattice geometries belong to the mean-field universality
class~\cite{ctmrg_54,ctmrg_pq}. If, however, a multistate spin system results in a
discontinuous (first-order) phase transition in the bulk, the phase-transition point
depends on the magnetic field $h_b$. In particular, if $h_b$ is varied on the outermost
boundary layer, the phase-transition temperature $T_{\rm pt}(h_b)$ changes continuously
within the temperature region $\Delta T>0$ for any $Q\geq3$ Potts models on the hyperbolic
geometries $\{p,q\}$.

With respect to our recent studies of quantum Ising and Heisenberg models on
hyperbolic lattices~\cite{TPVF54,TPVFp4}, which exhibit analogous physical properties
to those of the classical spins model, we expect the presence of identical features.
To prove this conjecture, we intend to perform extensive numerical studies
on quantum systems. The multistate Potts models on non-Euclidean lattices
were studied in this paper for their capability to mimic realistic interacting systems,
such as neural systems, social Internet networks, etc., in which the boundary
influences the behavior in the central part. The current remarkable physical feature
opens leads to an alternative view of the importance of boundary effects, as we have
shown that one can continuously control the thermodynamic properties of certain types
of complex systems on hyperbolic geometries.

A condensed-matter view point on the AdS-CFT correspondence is known to restrict
a negatively curved space geometry (AdS) to a preferred coordinate system, i.e.,
a lattice~\cite{Anderson}. We reduced the problem to an infinite set of
two-dimensional curved hyperbolic surfaces, where the underlying lattice geometry
$\{p,q\}$ varies by changing the two integer lattice parameters $p$ and $q$. The
impact of the $\{p,q\}$ geometry on the phase transitions with respect to the boundary
effects has been studied. Making use of the free-energy language, the boundary
structure of the hyperbolic geometry is naturally incorporated into the solution to
carry the essential feature of AdS space. Our intention for future studies
is to calculate the von Neumann entanglement entropy of a subsystem, e.g. a bulk,
for the quantum Heisenberg and transverse field Ising models on $\{4,q>3\}$ lattice
geometries~\cite{Daniska}. The impetus for this calculation lies in the concept of
holographic entanglement entropy~\cite{tHooft,Susskind,Takayanagi}, in which a
nongravitational theory is expected to exist on the boundary of the (bulk) subsystem
of ($d + 1$)-dimensional hyperbolic spaces. The entanglement entropy, which is related
to a reduced density matrix of the bulk subsystem, provides an appropriate measure
of the amount of information within the AdS-CFT correspondence. Hence, the entropy
is proportional to the boundary (minimal area surface) within AdS space, which links
the duality with the corresponding $d$-dimensional bulk region defined in CFT.

\begin{acknowledgments}
This work was supported by Vedeck\'{a} Grantov\'{a} Agent\'{u}ra M\v{S}VVa\v{S} SR a SAV (VEGA-2/0130/15) and Agent\'{u}ra na Podporu V\'{y}skumu a V\'{y}voja (QIMABOS-APVV-0808-12).
\end{acknowledgments}

\end{document}